\documentclass[journal]{IEEEtran}
\ifCLASSINFOpdf
\else
\fi 
\newtheorem{lemma}{Lemma}

\newtheorem{remark}{Remark}
\usepackage{amsfonts}
\usepackage{algorithmic}
\usepackage{algorithm}
\usepackage{tabulary}
\usepackage{tikz}
\usepackage{array}
\usepackage[caption=false,font=normalsize,labelfont=scriptsize,textfont=scriptsize]{subfig}
\usepackage{textcomp}
\usepackage{stfloats}
\usepackage{bm}
\usepackage{multirow} 
\usepackage{url}
\usepackage{verbatim}
\usepackage{graphicx}
\usepackage{subfig}
\usepackage{cite}
\usepackage{color}
\usepackage{amssymb}
\usepackage{siunitx}
\usepackage{threeparttable}
\usepackage{amsmath}
\usepackage{amssymb}
\usepackage{graphicx} 
\usepackage{subfig}
\usepackage[export]{adjustbox}
\begin{document}
\title{Sparse Channel Estimation for SIM-based\\ mmWave Near-Field Communications}
\author{Xianghao~Yao,
 Jiancheng~An, {\emph{Senior Member, IEEE}}, Enyu Shi, Jiayi Zhang, {\emph{Senior Member, IEEE}},\\ Lu Gan, {\emph{Member, IEEE}}, Michail Matthaiou, {\emph{Fellow, IEEE}},\\ Symeon Chatzinotas, {\emph{Fellow, IEEE}}, and Marco Di Renzo, {\emph{Fellow, IEEE} } 
\thanks{This work was supported by the National Natural Science Foundation of China (NSFC) under Grant 62471096. The work of M. Matthaiou was supported by the European Research Council (ERC) under the European Union’s Horizon 2020 research and innovation programme (grant agreement No. 101001331). This paper has been presented in part at the IEEE International Conference on Communications (ICC), June 2025 \cite{Yao2025ICC}. \emph{(Corresponding Author: Jiancheng An.)}}
\thanks{X. Yao and L. Gan are with the School of Information and Communication Engineering, University of Electronic Science and Technology of China (UESTC), Chengdu, Sichuan 611731, China, and also with the Yibin Institute of UESTC, Yibin 644000, China (e-mail: xianghao\_yao@163.com; ganlu@uestc.edu.cn).}
\thanks{J. An is with the School of Electronic Science and Engineering, University of Electronic Science and Technology of China (UESTC), Chengdu, 611731, China (e-mail: jiancheng.an@uestc.edu.cn).}
\thanks{E. Shi and J. Zhang are with the State Key Laboratory of Advanced Rail Autonomous Operation, and also with the School of Electronics and Information Engineering, Beijing Jiaotong University, Beijing 100044, P. R. China (e-mail: {enyushi, jiayizhang}@bjtu.edu.cn).}
\thanks{M. Matthaiou is with the Centre for Wireless Innovation (CWI), Queen’s University Belfast, BT3 9DT Belfast, U.K. (email: m.matthaiou@qub.ac.uk).}
\thanks{S. Chatzinotas is with SnT, University of Luxembourg, 1855 Luxembourg City, Luxembourg, and with College of Electronics \& Information, Kyung Hee University, Yongin-si, 17104, Republic of Korea (e-mail: symeon.chatzinotas@uni.lu).}
\thanks{M. Di Renzo is with CNRS and CentraleSup\'elec, Institute of Electronics and Digital Technologies (IETR), Avenue de la Boulaie, 35576 Cesson-S\'evign\'e, France (marco.direnzo@centralesupelec.fr), and with King's College London, Department of Engineering - Centre for Telecommunications Research, WC2R 2LS London, United Kingdom (marco.di\_renzo@kcl.ac.uk).}
\thanks{The work of M. Di Renzo was supported in part by the European Research Council (ERC) under the European Union’s Horizon Europe Programme WePhICom (agreement number 101225119), as well as by the European Union through the Horizon Europe project COVER under grant agreement number 101086228, the Horizon Europe project UNITE under grant agreement number 101129618, the Horizon Europe project INSTINCT under grant agreement number 101139161, and the Horizon Europe project TWIN6G under grant agreement number 101182794, as well as by the Agence Nationale de la Recherche (ANR) through the France 2030 project ANR-PEPR Networks of the Future under grant agreement NF-PERSEUS 22-PEFT-004, and by the CHIST-ERA project PASSIONATE under grant agreements CHIST-ERA-22-WAI-04 and ANR-23-CHR4-0003-01. Also, the work of M. Di Renzo was supported in part by the Engineering and Physical Sciences Research Council (EPSRC), part of UK Research and Innovation, and the UK Department of Science, Innovation and Technology through the CHEDDAR Telecom Hub under grant EP/Y037421/1, through the HASC Telecom Hub under grant EP/Y037197/1, and through the TITAN Telecom Hub under grant EP/Y037243/1.}
}
\maketitle
\markboth{DRAFT}%
{Shell \MakeLowercase{\textit{et al.}}: Bare Demo of IEEEtran.cls for IEEE Journals}
\begin{abstract}
Accurate acquisition of channel state information (CSI) is essential for fully harnessing the potential of stacked intelligent metasurfaces (SIMs) in communication systems. In this paper, we address the channel estimation (CE) problem in SIM-based multi-user (MU) millimeter-wave (mmWave) near-field communication systems. To address the severe path loss and blockage in mmWave communication systems, many meta-atoms are typically integrated into each layer of the SIM. Then, the number of radio frequency (RF) chains at the base station (BS) is fewer than that of meta-atoms per layer, resulting in an underdetermined problem. Additionally, the increase in the number of meta-atoms in each layer expands the SIM's near-field region, leading to the user equipment (UEs) being mostly situated in this region, necessitating precise modeling of the channel under the spherical wavefront assumption. To address these issues, we introduce a compressed sensing (CS)-based CE protocol to tackle the underdetermined problem. In contrast to the traditional CS-based estimation framework, we investigate a polar-domain channel representation to tackle the severe energy spread effect of the classical angular-domain channel representation in near-field communication systems. Specifically, we design a novel polar-domain transform matrix for uniform planar arrays (UPAs), thereby transforming the CE problem into a sparse recovery task of the paths' support set and complex gains. To overcome the limitations of the sparse Bayesian learning (SBL) framework in tackling high-dimensional dictionaries, we propose a low-complexity polar-domain SBL (LCPD-SBL) algorithm, which significantly reduces computational complexity without compromising estimation accuracy. Numerical simulation results demonstrate that the proposed polar-domain transform matrix yields a better estimation accuracy than traditional angular-domain approaches. Additionally, the proposed LCPD-SBL algorithm can be faster than existing SBL methods by up to $4\times$ while sustaining the same estimation performance. 
\end{abstract}
\IEEEpeerreviewmaketitle
\section{Introduction}
\label{Introduction}
\subsection{Background}
Millimeter-wave (mmWave) communication systems operating in the $\SI{30} - \SI{300}{GHz}$ range offer higher data transmission rates and lower latency \cite{MingMillimeter,li2017millimeter}. However, mmWave communication faces significant challenges due to high atmospheric attenuation from oxygen absorption and poor diffraction when encountering obstacles, resulting in severe path loss and blocking/outage phenomena \cite{akyildiz2018combating}. Fortunately, the smaller wavelength enables the transmitters and receivers to deploy more antenna elements, facilitating multiple-input multiple-output (MIMO) systems that combat path loss with beamforming gain \cite{MingMillimeter}. This, however, introduces complexity for hardware design and results in increased power consumption and cost for radio-frequency (RF) chains. This discussion underscores the pressing necessity for advanced transceiver technology with high energy efficiency \cite{Jiayi5G}.

To address these issues, the disruptive stacked intelligent metasurface (SIM) technology has emerged thanks to its fewer RF chains and low power consumption  \cite{an2024emerging,papazafeiropoulos2024achievable,papazafeiropoulos2024performance,papazafeiropoulos2024achievable2,Huang2025SIM,Niu2025SIM,Darsena2025SIM,Lin2025SIM,Shi2025SIM,An2025SIM,Liu2025SIM,Javed2025SIM,an2026stacked}. Specifically, an SIM consists of multiple stacked layers of programmable metasurfaces, each layer equipped with many low-cost nearly-passive meta-atoms \cite{an2023stackedMIMO,an2024SIM_MIMO,Niu2024SIM,Stefan2024SIM,Yao2024SIM}. Within the multi-layered structure of the SIM, electromagnetic (EM) waves transmitted through the preceding layer's meta-atoms serve as secondary point sources, irradiating all subsequent layers' meta-atoms, following the Huygens–Fresnel principle \cite{lin2018all}. Each meta-atom in an SIM can function as a reprogrammable artificial neuron with adjustable weights, resembling a fully connected artificial neural network, allowing the SIM to adjust the EM response of each meta-atom by tunning the bias voltage of the attached control circuit \cite{Huang2025SIM,Niu2025SIM,Darsena2025SIM,Lin2025SIM,Shi2025SIM,An2025SIM,Liu2025SIM,Javed2025SIM}. Additionally, by properly configuring each layer's meta-atom complex-valued transmission coefficients, the SIM can manipulate the EM waves propagating at the speed of light in a layer-by-layer manner enabling it to accomplish advanced signal processing tasks \cite{2024SIMAn,2025SIMHuai,2025SIMPei}.\footnote{ In recent studies, SIM prototypes have been shown to be able to implement MIMO precoding, direction of arrival (DoA) estimation, etc. within the wave domain \cite{liu2022programmable}.}

Leveraging the wave-domain beamforming capabilities, recent research has focused on integrating SIMs with transceivers, replacing expensive active components with programmable metasurfaces to achieve signal processing during the wireless propagation \cite{an2024emerging,papazafeiropoulos2024achievable,papazafeiropoulos2024performance,papazafeiropoulos2024achievable2,Huang2025SIM,Niu2025SIM,Darsena2025SIM,Lin2025SIM,Shi2025SIM,An2025SIM,Liu2025SIM,Javed2025SIM}. More specifically, An \emph{et al.} proposed an SIM-based holographic MIMO (HMIMO) framework \cite{an2023stackedMIMO}, where EM waves transmitted through the SIM automatically undergo transmit precoding and receive combining. This architectural design significantly reduces the number of RF chains \cite{an2023stackedMIMO}. By capitalizing on this advanced framework, Papazafeiropoulos \emph{et al.} proposed a projected gradient ascent (PGA) algorithm to simultaneously optimize the transmitted signal covariance matrix and the SIM's phase shifts to effectively maximize the achievable rate \cite{papazafeiropoulos2024achievable}. Furthermore, Shi \emph{et al.} applied SIM technology to a cell-free massive MIMO network, aiming to enhance both spectral and energy efficiency \cite{2025SIMShi}.\par
Although the advantages of SIM-based communication systems have been preliminarily validated via both theoretical analysis and experimental results, significant challenges remain in practical deployments \cite{nerini2024physically,wang2024multi,li2024stacked}. Specifically, precise channel state information (CSI) is indispensable for realizing wave-domain beamforming and desired signal processing functions. However, a large number of meta-atoms are typically integrated into each layer of the SIM to achieve effective wave-based beamforming and fully exploit array gains to mitigate path loss in mmWave communications \cite{2024SIMAn,an2023stackedMIMO}. This imposes two critical obstacles to channel estimation (CE) within SIM-based mmWave communication systems:
\begin{itemize}
 \item\textbf{Underdetermined Problem:}
Due to the passive nature of meta-atoms, SIMs generally lack the capability to actively sense or estimate the individual wireless channels associated with each meta-atom. As a result, the base station (BS) does not have prior knowledge of the effective channel subspace. Moreover, the large-scale deployment of meta-atoms leads to a substantial increase in the channel dimensionality. Given the limited number of RF chains available at the BS, CE for a large number of meta-atoms naturally results in an underdetermined estimation problem.
 \item\textbf{Expansion of Near-Field Region:}
 In SIM-aided systems, the increased number of meta-atoms in each layer results in an enlargement of the aperture and expands the near-field region \cite{ramezani2023near}. Hence, precise modeling of the spherical wavefronts is indispensable for integrating the SIM technology in mmWave communication systems.
\end{itemize}
\subsection{Related Works}
\begin{table*}[t]
\renewcommand\arraystretch{1.5}
\centering
\caption{Comparison Between the Proposed CE Scheme and Related Works}
\begin{tabulary}{\textwidth}{C|C|C|C|C|C|C}
\hline
\multicolumn{1}{c|}{Reference} & System & \begin{tabular}[c]{@{}c@{}}SIM\end{tabular} & \begin{tabular}[c]{@{}c@{}}Near/Far-field\end{tabular} & \begin{tabular}[c]{@{}c@{}}Transform\end{tabular} & \begin{tabular}[c]{@{}c@{}}Array\end{tabular} & Algorithms \\ \hline
Lee \emph{et al.} \cite{lee2016channel} & MIMO & $\times$ & Far-field & Angular domain &UPA & OMP 
 \\ \cline{1-1} \cline{2-7} 
Mishra \emph{et al.} \cite{mishra2017sparse} & MIMO & $\times$ & Far-field & Angular domain &UPA & SBL 
 \\ \cline{1-1} \cline{2-7} 
Cui \emph{et al.} \cite{cui2022channel} & MIMO & $\times$ & Near-field & Polar domain 
&ULA & P-SOMP, P-SIGW \\ \cline{1-1} \cline{2-7} 
Yao \emph{et al.} \cite{yao2024channel} & MU MISO & \checkmark & Far-field & $\times$ &UPA & Codebook \\
\cline{1-1} \cline{2-7}
Nadeem \emph{et al.} \cite{nadeem2023hybrid} & MU MISO & \checkmark & Far-field & $\times$ &UPA & Gradient descent \\ \cline{1-1} \cline{2-7} 
 This paper & MU MISO & \checkmark & Near-field & Polar domain &UPA & LCPD-SBL 
\\ \hline 
\multicolumn{7}{l}
{{\scriptsize P-SOMP: Polar-domain simultaneous orthogonal matching pursuit algorithm.
P-SIGW: Polar-domain simultaneous iterative gridless weighted algorithm.}}
\end{tabulary}
\end{table*}
Some recent studies have proposed various solutions to partially tackle the aforementioned challenges \cite{yao2024channel,nadeem2023hybrid}. For instance, Yao \emph{et al.} proposed a CE protocol for multi-user (MU) multiple-input single-output (MISO) systems, where different user equipment (UEs) transmit mutually orthogonal pilot sequences to the BS over multiple pilot blocks \cite{yao2024channel}. In this approach, the BS alleviates the underdetermined nature of the CE problem by collecting multiple replicas of the pilot signals, thereby effectively increasing the number of observations. However, the number of required pilot blocks scales linearly with the number of meta-atoms in the last SIM layer, which leads to excessive pilot overhead and limits the scalability of the conventional CE scheme in large-scale SIM-assisted systems. Additionally, Nadeem \emph{et al.} developed a hybrid digital-wave domain CE framework, which first processes the received training symbols in the wave domain by configuring the SIM, followed by low-complexity digital-domain processing \cite{nadeem2023hybrid}. The wave-domain estimator is optimized via a gradient descent algorithm that iteratively minimizes the channel estimator's mean squared error (MSE). \par
Notably, the linear CE approaches in \cite{yao2024channel,nadeem2023hybrid} still incur substantial pilot overhead. In conventional mmWave communication systems without SIMs, to reduce the pilot overhead, several CE schemes incorporating compressed sensing (CS) techniques have been developed \cite{lee2016channel,mishra2017sparse,cui2022channel}. Specifically, Lee \emph{et al.} exploited the channel sparsity in the angular domain by establishing a parameterized channel model with quantized angles of departure/arrival (AoDs/AoAs). This approach converts the CE problem into a sparse signal recovery task. To solve this task, an orthogonal matching pursuit (OMP) algorithm was applied to efficiently estimate the AoDs/AoAs and their corresponding gains in \cite{lee2016channel}. However, this approach assumes that AoDs/AoAs are discrete (on-grid), and thus the accuracy is limited by the grid resolution. To address this issue, Mishra \emph{et al.} conceived a sparse Bayesian learning (SBL)-based CE framework for hybrid mmWave MIMO systems, which achieves significant performance gains over OMP in both on-grid and practical scenarios \cite{mishra2017sparse}.

However, existing low-overhead CE schemes rely heavily on the channel sparsity in the angular domain \cite{lee2016channel,mishra2017sparse,cui2022channel}. Such sparsity may be unachievable in SIM-based near-field communication systems, as the classical angular-domain channel representation introduces severe energy spread effect \cite{cui2022channel}. Specifically, the energy of a single near-field path component spreads across multiple angles, causing the angular-domain representation of the near-field channel to no longer be sparse. To deal with this issue, Cui \emph{et al.} \cite{cui2022channel} designed a polar-domain transform matrix to convert the near-field channel into a polar-domain representation that simultaneously considers angular and distance dimensions, ensuring the near-field channel sparsity in the polar domain. However, the designed polar-domain transform matrix is only applicable to uniform linear arrays (ULAs). 
\subsection{Motivation and Contributions}
In a nutshell, existing CE schemes cannot be directly implemented in our system due to the following reasons:
\begin{enumerate}
 \item Current CE schemes for SIM-based communication systems typically lead to significant pilot overhead.
 \item Most low-overhead CE methods in mmWave communication systems rely on the channel sparsity in the angular domain. Although \cite{cui2022channel} accounted for the energy spread effect in the near field, the polar-domain transform matrix proposed was designed for ULAs, whereas SIMs are typically implemented by uniform planar arrays (UPAs).
 \item Due to the large-scale deployment of meta-atoms on each SIM layer, the channel in our system is inherently high-dimensional. Therefore, conventional SBL-based CE schemes, which involve high-dimensional matrix inversion and storage, are computationally intensive and impractical.
\end{enumerate}\par
To address these issues, we design a novel polar-domain transform matrix for UPAs to transform the CE problem into a sparse recovery task of the paths’ support set and corresponding complex gains. To elaborate, we have summarized the comparison between our contributions and related works in Table I. The specific contributions of our work are outlined below:
\begin{enumerate}
 \item We introduce a CE protocol for SIM-based uplink mmWave near-field communication systems. To mitigate MU interference, we employ mutually orthogonal pilot sequences sent from multiple UEs to the BS. Furthermore, we leverage the CS framework to effectively tackle the underdetermined problem. Multiple successive blocks of pilot signals are acquired for enhancing uplink CE performance.
 \item We develop a polar-domain representation of the near-field channel, thereby transforming the CE problem into a sparse recovery task of the paths’ support set and complex gains. Additionally, to minimize the column coherence of the polar-domain transform matrix, we propose effective angular and distance sampling methods to construct a polar-domain transform matrix for the UPAs.
 \item The SBL framework is employed to solve the sparse signal recovery task. To alleviate the computational burden of the SBL's inference process, we propose a low-complexity polar-domain SBL (LCPD-SBL) algorithm by introducing a cost-efficient covariance-free expectation-maximization (CoFEM) algorithm to replace the predominant expectation-maximization algorithm in the SBL's inference process. 
 \item Finally, numerical results are presented to demonstrate the performance superiority of the CE methods based on the proposed polar-domain transform matrix over the existing angular-domain approaches. Moreover, the proposed LCPD-SBL algorithm can be faster than existing SBL methods by up to $4\times$ while achieving the same normalized mean squared error (NMSE) performance.
\end{enumerate}
\begin{table}[!t]
\renewcommand\arraystretch{1.5}
 \begin{center} 
 \caption{Major Symbols and Their Definitions}
 \label{tab:table3}
 \begin{tabular}{c|l} 
 \hline
 Symbol & Meaning \\
 \hline\hline
 $L/K/Q$ & Number of SIM layers / UEs / propagation paths \\
 \hline
 $\lambda/\sigma_n^2/\kappa$ & Wavelength / Noise power / Control parameter \\
 \hline
 $M/N$ & Number of BS antennas / meta-atoms on each layer \\
 \hline
 $N_{H}/N_{V}$ & Number of meta-atoms per row / column\\
 \hline
 $\Psi$ & Number of pilot blocks \\
 \hline
 $\tau_{p}$ & Length of pilot sequences\\
 \hline
 $\Delta$ & Distance between adjacent meta-atoms \\
 \hline
 $p_{k}$ & Average power of the $k$-th UE's pilot signals  \\
 \hline
 $N_{G}$ & Number of sampled near-field array response vectors \\
 \hline
 $\mathbf{s}_{k}/\bm{\Pi }$ & Pilot sequence sent by the $k$-th UE / Sensing matrix\\
 \hline
 $\mathbf{b}\left( \cdot \right )$ & Near-field array response vector \\
 \hline
 $\mathbf{B}$ & Proposed polar-domain transform matrix \\
 \hline
 $\bm{\Phi}^{l}$ & Transmission coefficient matrix for the $l$-th SIM layer \\
 \hline
 $\mathbf{G}$ & SIM-based wave-domain beamforming matrix \\
 \hline
 $\mathbf{h}_{k}^H$ & Propagation channel from the last SIM layer to the $k$-th UE \\
 \hline
 \end{tabular}
 \end{center}
\end{table}
\subsection{Organization and Notations}
The remainder of this paper is structured as follows: In Section \ref{SYSTEM}, the SIM-based mmWave near-field communication system, channel model, and CE protocol are introduced. In Section \ref{polar_domain}, the polar-domain representation of the near-field channel and the design method of the polar-domain transform matrix are given. Section \ref{CS_scheme} presents the LCPD-SBL algorithm. We provide numerical results in Section \ref{simulation}, while Section \ref{Conclusion} summarizes the paper’s findings and concludes the work. \par
\textit{Notations}: Table \ref{tab:table3} summarizes the key notations and their definitions used in this paper. Scalar variables are denoted by lower-case letters (e.g., $x$), column vectors by bold lower-case letters (e.g., $\mathbf{x}$), and matrices by bold upper-case letters (e.g., $\mathbf{X}$). For any vector $\mathbf{x}$, its $\ell_p$-norm takes the form $\| \mathbf{x} \|_{p}$. For any matrix $\mathbf{X}$, its transpose, conjugate transpose, and inverse are denoted as $\mathbf{X}^T$, $\mathbf{X}^H$, and $\mathbf{X}^{-1}$, respectively; $\mathbf{I}_{N}$ denotes the $N \times N$ identity matrix; $\mathbb{C}^{M\times N}$ denotes the space of all complex $M \times N$ matrices. Vertical concatenation of matrices $\mathbf{X}$ and $\mathbf{Y}$ is denoted by $\left [ \mathbf{X};\mathbf{Y} \right ]$, while horizontal concatenation is denoted by $\left [ \mathbf{X},\mathbf{Y} \right ]$. A complex Gaussian random variable with mean $\bm{\mu}$ and covariance $\bm{\Sigma}$ is denoted by $\mathcal{CN}(\bm{\mu}, \bm{\Sigma})$, whereas a uniform distribution between $a$ and $b$ is denoted by $\mathcal{U}(a, b)$. The expectation operator is denoted by $\mathbb{E}\left \{ \cdot \right \}$, and the operator $\text{diag}(\mathbf{a})$ denotes the construction of a diagonal matrix from the vector $\mathbf{a}$. The absolute value of a scalar is denoted by $\left | \cdot \right |$, and the symbol $\sim$ indicates “distributed as”. Finally, $\left \lceil x \right \rceil$ is defined as the smallest integer greater than or equal to $x$.
\section{System and Channel Models}
\label{SYSTEM}
In this section, we introduce the system model, channel model, and CE protocol for an SIM-based uplink mmWave near-field communication system.
\subsection{System Model}
\begin{figure}[t]
\centering 
\includegraphics[width=8.5cm]{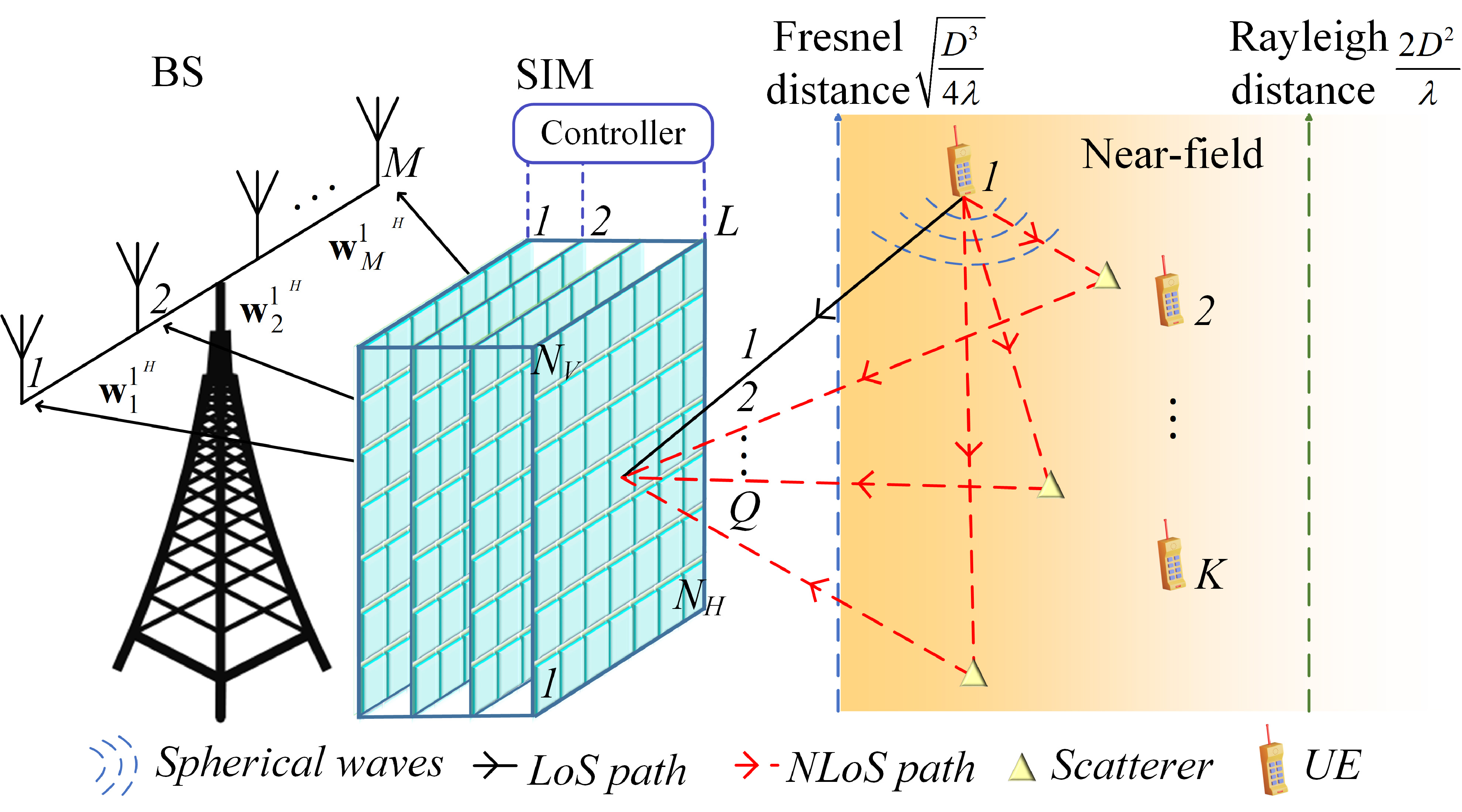} 
\caption{The SIM-based MU mmWave near-field communication system.}
\label{System}
\end{figure}

As illustrated in Fig. \ref{System}, a BS equipped with a ULA having $M$ antennas serves $K$ single-antenna UEs. The SIM is integrated with the radome of the BS and connected to an intelligent controller, such as a field programmable gate array (FPGA) board. This configuration enables the SIM to impose a separate and adjustable phase shift to the EM waves propagated through each meta-atom. Thanks to the multi-layer structure, the SIM can perform advanced signal processing tasks directly within the wave domain \cite{2024SIMAn}. Specifically, the SIM consists of $L$ metasurface layers, with each layer modeled as a UPA containing $N$ meta-atoms, where we have $N\gg M$ \cite{2024SIMAn}. The number of meta-atoms in each row and each column of the SIM in each layer is denoted as $N_{H}$ and $N_{V}$, respectively, resulting in $N = N_{H} \times N_{V}$. Also, $ \Delta$ denotes the distance between adjacent meta-atoms, while each meta-atom has a length of $d_{x}$ and a width of $d_{y}$. The index sets $\mathcal{L} = \left \{1,2,\dots, L \right \}$, $\mathcal{M} =\left \{ 1,2,\dots, M \right \} $, $\mathcal{N} =\left \{1,2,\dots, N \right \}$ and $\mathcal{K} =\left \{1,2,\dots, K \right \}$ are defined to correspond to the metasurface layers, antennas at the BS, meta-atoms on each SIM layer, and UEs, respectively. Additionally, $\check{r}_{n,m}$ denotes the spacing between the $m$-th antenna and the $n$-th meta-atom of the first SIM layer; $d_{t}$ denotes the horizontal distance between the antenna array and the first SIM layer; $d_{\text{Layer}}$ denotes the spacing between two adjacent SIM layers. \par
The transmission coefficient of the $n$-th meta-atom on the $l$-th SIM layer is denoted by $e^{j\theta _{n}^{l} }$, where $\theta _{n}^{l}\in \left [0,2\pi \right )$. As shown in Fig. \ref{System}, we assume that all layers of the SIM are parallel to each other, and the centers of all metasurface layers are in the same horizontal line in this paper. Furthermore, the transmit antenna array is arranged in parallel with the SIM, with all UEs positioned within the near-field region \cite{you2023near}.\par 
According to \cite{2024SIMAn}, the transmission coefficient matrix for the $l$-th SIM layer is given by 
\begin{equation}
\label{Phi_l}
 \mathbf{\Phi}^{l}=\text{diag}\left ( e^{j\theta _{1}^{l}},e^{j\theta _{2}^{l}},\dots ,e^{j\theta _{N}^{l}} \right )\in \mathbb{C}^{N\times N}.
\end{equation}
Moreover, $\mathbf{W}^{l} \in \mathbb{C}^{N\times N}, \forall l\ne 1, l\in\mathcal{L}$ denotes the transmission matrix from the $\left ( l-1 \right ) $-th to the $l$-th SIM layer. Based on Rayleigh-Sommerfeld's diffraction theory \cite{All}, we obtain the $(n,n')$-th element of $\mathbf{W}^{l}$ is
\begin{equation}
\label{W_l}
 w^{l}_{n,n'}= \frac{d_{x}d_{y}d_{\text{Layer} }}{\left(d^{l}_{n,n'}\right)^2} \left(\frac{1}{2\pi d^{l}_{n,n'}}-\frac{1}{\lambda }j \right)e^{j2\pi d^{l}_{n,n'}/\lambda }, 
\end{equation}
where $d^{l}_{n,n'}$ denotes the propagation distance from the $n'$-th meta-atom on the $\left ( l-1 \right )$-th SIM layer to the $n$-th meta-atom on the $l$-th SIM layer. Additionally, the transmission vector $\mathbf{w}_{m}^{1} \in \mathbb{C}^{N\times 1}$ from the $m$-th antenna to the first SIM layer can be calculated by replacing $d^{l}_{n,n'}$ and $d_\text{Layer}$ in (\ref{W_l}) with $\check{r}_{n,m}$ and $d_t$.\par 
Consequently, the SIM-based wave-domain beamforming matrix can be written as follows
\begin{equation}
 \mathbf{G}=\mathbf{\Phi} ^{L} \mathbf{W}^{L}\cdots \mathbf{\Phi} ^{2}\mathbf{W}^{2} \mathbf{\Phi} ^{1} \in \mathbb{C}^{N\times N}.
\end{equation}
\begin{remark}
 We assume that the transmission coefficients among neighboring SIM layers are ideal. In reality, deviations in the inter-layer transmission coefficients may exist due to the manufacturing precision limitations of metamaterial elements and deformations during assembly \cite{liu2024stacked}. Consideration of joint inter-layer transmission coefficient calibration and CE will be pursued in our future research.
\end{remark}
\subsection{Propagation Channel Model}
As a reasonable case study, we adopt a narrowband quasi-static flat fading channel model \cite{tse2005fundamentals}. Since the mmWave channel generally has a restricted number of scatterers, we adopt a clustered channel model comprising $Q$ scattering clusters, where each cluster has a propagation path between the UE and the SIM. Therefore, the mmWave channel from the $k$-th UE to the last SIM layer can be expressed as \cite{li2017millimeter,alkhateeb2014channel,el2014spatially}
\begin{equation}
\label{channel1}
 \mathbf{h}_{k}=\sqrt{\frac{N}{Q} } \sum_{q=1}^{Q}\beta _{q,k}\mathbf{b}\left ( \vartheta _{q,k},\gamma _{q,k},r _{q,k} \right ), 
\end{equation}
where $\beta_{q,k}$ is the complex gain corresponding to the $q$-th path of the $k$-th UE, $\vartheta _{q,k}\in\left [ -\frac{\pi}{2},\frac{\pi}{2} \right ],\gamma _{q,k} \in\left [ -\frac{\pi}{2},\frac{\pi}{2} \right ] $ are the associated azimuth and elevation AoAs, and $r _{q,k}$ is the distance between the UE (scatterer) and the center of the last SIM layer associated with the $q$-th path of the $k$-th UE.

Unlike the far-field array response vector derived under the planar wave assumption, the near-field array response vector is formulated based on the more accurate spherical wave assumption, expressed as \cite{cui2022channel}
\begin{equation}
\begin{aligned}
\label{b}
 \mathbf{b}&\left ( \vartheta _{q,k},\gamma _{q,k},r _{q,k} \right )\\&=\frac{1}{\sqrt{N}}\left [e^{-j\frac{2\pi}{\lambda }\left ( r_{q,k}^{1}-r_{q,k} \right ) };\dots;e^{-j\frac{2\pi}{\lambda }\left ( r_{q,k}^{N}-r_{q,k} \right ) } \right ], 
\end{aligned}
\end{equation}
where $r_{q,k}^{n}$ denotes the propagation distance between the $k$-th UE (scatterer) and the $n$-th meta-atom on the last SIM layer associated with the $q$-th path. Note that $r_{q,k}^n$ can be derived from $ \vartheta _{q,k},\gamma _{q,k}$, and $r_{q,k}$, where the details will be provided in Section III-B.
\begin{figure}[t]
\centering 
\includegraphics[width=8.5cm]{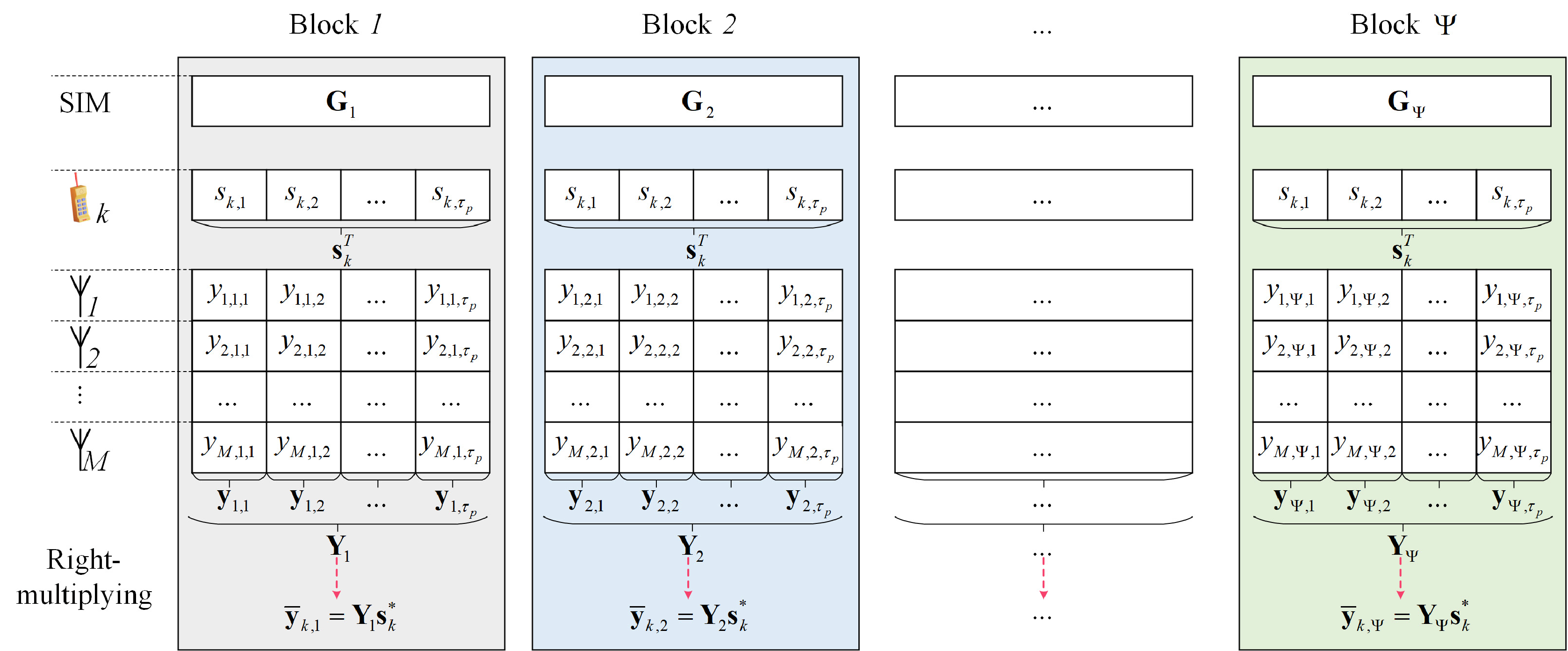} 
\caption{The proposed CE protocol.}
\label{protocol}
\end{figure}
\subsection{CE Protocol}
We address uplink CE in SIM-based mmWave communication systems in this paper. As depicted in Fig. \ref{protocol}, the estimation stage involves multiple blocks. In each block, $K$ UEs send mutually orthogonal pilot sequences of length $\tau_{p}$ to the BS. The pilot signal transmitted by the $k$-th UE during the $t$-th time slot $s_{k,t}$ satisfies 
\begin{equation}
\sum_{t = 1}^{\tau_{p}}s_{k,t}s_{k',t}^{*} = 
\begin{cases}
\tau_{p}, & \text{if } k = k', \\
0, & \text{if } k \neq k'.
\end{cases}
\end{equation}
It is worth noting that to generate $K$ orthogonal sequences, $\mathbf{s}_{k}$'s length must satisfy $\tau_{p}\ge K$. In this paper, to minimize the orthogonal pilot overhead, we set $\tau_{p}= K$ \cite{yao2024channel}.\par
As shown in Fig. \ref{protocol}, during the $t$-th time slot of the $\psi$-th block, the signal received by the $m$-th antenna can be expressed as
\begin{equation}
\label{signal}
 \begin{aligned}
 y_{m,\psi,t}=\left( \mathbf{w}_{m}^1\right)^{H}\mathbf{G}_{\psi}^{H}\sum_{k'=1}^{K}\mathbf{h}_{k'} \sqrt{p_{k'}}s_{k',t}+n_{m,\psi,t},
 \end{aligned}
\end{equation}
where $p_{k}$ denotes the average power of the $k$-th UE's pilot signal, $\mathbf{G}_{\psi} \in \mathbb{C}^{N \times N}$ denotes the SIM-based wave-domain beamforming matrix in the $\psi$-th block, $n_{m,\psi, t}\sim \mathcal{CN}\left (0,\sigma _{n}^2 \right )$ denotes the noise received at the $m$-th antenna during the $t$-th time slot of the $\psi$-th block, while $\sigma _{n}^2$ is the noise power. Upon collecting the output signals associated with $M$ antennas during the $t$-th time slot of the $\psi$-th block, we arrive at
\begin{equation}
\mathbf{y}_{\psi,t}=\mathbf{W}^{H}\mathbf{G}_{\psi}^{H}\sum_{k'=1}^{K}\mathbf{h}_{k'} \sqrt{p_{k'}}s_{k',t}+\mathbf{n}_{\psi,t},
\end{equation}
where $\mathbf{y}_{\psi,t}=\left [ y_{1,\psi,t};y_{2,\psi,t};\dots;y_{M,\psi,t} \right ]\in \mathbb{C}^{M\times 1}$ denotes the received signal, $\mathbf{W}=\left [ \mathbf{w}_{1}^1,\mathbf{w}_{2}^1,\dots,\mathbf{w}_{M}^1 \right ] \in \mathbb{C}^{N \times M}$ denotes the transmission matrix from the BS to the first SIM layer, and $\mathbf{n}_{\psi,t}=\left [ n_{1,\psi,t};n_{2,\psi,t};\dots;n_{M,\psi,t}\right ] \in \mathbb{C}^{M \times 1}$ denotes the noise vector, during the $t$-th time slot of the $\psi$-th block, respectively.\par

After collecting the received signals over $\tau _{p}$ time slots in the $\psi$-th block, we have
\begin{equation}
\label{multi_singal}
 \mathbf{Y_{\psi}}=\mathbf{W}^{H}\mathbf{G}_{\psi}^{H}\sum_{k'=1}^{K}\mathbf{h}_{k'} \sqrt{p_{k'}}\mathbf{s}_{k'}^{T}+\mathbf{N_{\psi}},
\end{equation}
where $\mathbf{Y_{\psi}}=\left [ \mathbf{y}_{\psi,1},\mathbf{y}_{\psi,2},\dots,\mathbf{y}_{\psi,\tau_{p}} \right ] \in \mathbb{C}^{M \times \tau_{p}} $ denotes the received signal, $\mathbf{s}_{k}=\left [ s_{k,1};s_{k,2};\dots;s_{k,\tau_{p}} \right ] \in \mathbb{C}^{\tau_{p} \times 1}$ denotes the pilot sequence sent by the $k$-th UE, and $\mathbf{N_{\psi}}=\left [ \mathbf{n}_{\psi,1},\mathbf{n}_{\psi,2},\dots,\mathbf{n}_{\psi,\tau_{p}} \right ] \in \mathbb{C}^{M \times \tau_{p}}$ denotes the noise in $\tau_{p}$ time slots of the $\psi$-th block, respectively. Next, the $k$-th UE's signal in the $\psi$-th block can be obtained by right-multiplying $\mathbf{s}_{k}^*$ on both sides of ($\ref{multi_singal}$), yielding
\begin{equation}
\label{singal}
 \mathbf{\Bar{y}}_{k,\psi}=\sqrt{p_{k}}\tau_{p}\mathbf{W}^{H}\mathbf{G}_{\psi}^{H}\mathbf{h}_{k} +\mathbf{N}_{\psi}\mathbf{s}_{k}^*,
\end{equation}
where $\mathbf{\Bar{y}}_{k,\psi}=\mathbf{Y}_{\psi}\mathbf{s}_{k}^* \in \mathbb{C}^{M \times 1}$ denotes the received signal vector of the $k$-th UE within the $\psi$-th block, corresponding to $M$ antennas.\par 
Note that the dimension of the channel ($N$) is higher than that of the observation signal ($M$), thereby creating an underdetermined problem. Hence, directly applying the conventional CE methods often results in substantial pilot overhead \cite{yao2024channel,nadeem2023hybrid}. To tackle this issue, we consider leveraging the CS technique to effectively reduce the pilot overhead. Additionally, to enhance the CE performance, we stack observations over $\Psi$ blocks, yielding
\begin{equation}
\label{CS2}
\tilde{\mathbf{y}}_{k} =\sqrt{p_{k}}\tau_{p}
\mathbf{P}\mathbf{h}_{k}+
\tilde{\mathbf{n}}_{k}, 
\end{equation}
where $\tilde{\mathbf{y}}_{k}=\left [ \mathbf{\Bar{y}}_{k,1};\mathbf{\Bar{y}}_{k,2};\dots;\mathbf{\Bar{y}}_{k,\Psi } \right ] \in \mathbb{C}^{\Psi M \times 1}$ denotes the received signal vector of the $k$-th UE, $ \mathbf{P}=\left [\mathbf{W}^{H}\mathbf{G}_{1}^{H};\mathbf{W}^{H}\mathbf{G}_{2}^{H};\dots;\mathbf{W}^{H}\mathbf{G}_{\Psi}^{H}\right ] \in \mathbb{C}^{\Psi M\times N}$ denotes the transmission matrix from the last SIM layer to the BS, and $\tilde{\mathbf{n}}_{k}=\left [ \mathbf{N}_{1}\mathbf{s}_{k}^*;\mathbf{N}_{2}\mathbf{s}_{k}^*;\dots;\mathbf{N}_{\Psi}\mathbf{s}_{k}^* \right ]\in\mathbb{C}^{\Psi M\times1}$ denotes the noise, over $\Psi$ successive blocks, respectively. In the next section, we will introduce the polar-domain representation of the channel $\mathbf{h}_k$, leveraging the channel sparsity in the polar domain to reformulate and solve the CE problem in (\ref{CS2}).\par
\section{Polar-Domain Representation for SIM-based Communication Systems}
\label{polar_domain}
In this section, we investigate the polar-domain representation of the mmWave channel between the UE and the last layer of the SIM. We aim to construct a polar-domain transform matrix that can approximately represent the near-field channels of all UEs without requiring prior knowledge of their exact propagation parameters. To this end, we further propose angular and distance sampling methods to generate the polar-domain transform matrix for UPAs that yields the optimal estimation performance.
\subsection{Polar-Domain Representation of the Near-Field Channel}
Specifically, the near-field channel defined in (\ref{channel1}) can be expressed in a more concise form as \cite{el2014spatially}
\begin{equation}
\label{channel2}
 \mathbf{h}_{k}=\mathbf{B}_{k} \bm{\beta}_{k} ,
\end{equation}
where $\bm{\beta}_{k}=\sqrt{\frac{N}{Q} } \left [ \beta_{1,k};\beta_{2,k};\dots;\beta_{Q,k} \right ]\in \mathbb{C}^{Q \times 1}$, $\mathbf{B}_{k}= \left [ \mathbf{b}\left ( \vartheta _{1,k},\gamma _{1,k},r _{1,k} \right ),\dots,\mathbf{b}\left ( \vartheta _{Q,k},\gamma _{Q,k},r _{Q,k} \right ) \right ]\in \mathbb{C}^{N \times Q}$ represents the transform matrix that contains the near-field array response vectors corresponding to the last SIM layer.\par 
From (\ref{b}), it can be observed that the phase of each element in the near-field array response vector is nonlinear to the meta-atom index $n$. Therefore, we cannot directly convert the channel to its angular domain via a Fourier transform matrix to obtain a sparse representation of $\mathbf{h}_k$ \cite{cui2022channel}. Taking this into account, next we aim to design a uniform polar-domain transform matrix $\mathbf{B}= \left [ \mathbf{b}\left ( \bar{\vartheta} _{1},\bar{\gamma} _{1},\bar{r} _{1} \right ),\dots,\mathbf{b}\left ( \bar{\vartheta} _{N_{G}},\bar{\gamma} _{N_{G}},\bar{r} _{N_{G}} \right ) \right ]\in \mathbb{C}^{N \times N_{G}}$ for the UPA to transform the channel into its polar-domain representation, where $N_{G}$ is the number of the sampled near-field array response vectors, while $\bar{\vartheta} _{q}$, $\bar{\gamma} _{q}$, and $\bar{r}_{q}$ are the corresponding channel parameters at the sampled points.\par 
It can be observed that when $ N_G $ is sufficiently large, we can reasonably assume that the polar-domain transform matrix $ \mathbf{B} $ is capable of covering all the actual near-field array response vectors present in the mmWave channels of the $ K $ UEs.
Therefore, the polar-domain representation of the near-field channel between the $k$-th UE and the last SIM layer can be given by
\begin{equation}
 \label{channel3}
 \mathbf{h}_{k}=\mathbf{B}\bm{\bar{\beta}}_{k} ,
\end{equation}
where $\bm{\bar{\beta}}_{k}=\sqrt{\frac{N}{Q} } \left [ \bar{\beta}_{1,k};\bar{\beta}_{2,k};\dots;\bar{\beta}_{N_{G},k} \right ] \in \mathbb{
C}^{N_{G} \times 1}$, with each of its entries denoting the complex gain associated with the corresponding path of the $k$-th UE. Therefore, $\bm{\bar{\beta}}_{k}$ is a sparse vector with $Q$ non-zeros entries. Substituting (\ref{channel3}) into (\ref{CS2}), we have
\begin{equation}
\label{CS}
\begin{aligned}
 \tilde{\mathbf{y}}_{k}&=\sqrt{p_{k}}\tau_{p}\mathbf{P}\mathbf{B}\bm{\bar{\beta}}_{k} +\tilde{\mathbf{n}}_{k}\\
&=\bm{\Pi }\bm{\bar{\beta}}_{k}+\tilde{\mathbf{n}}_{k},
\end{aligned}
\end{equation}
where $\bm{\Pi }=\sqrt{p_{k}}\tau_{p}\mathbf{P}\mathbf{B} \in \mathbb{C}^{\Psi M \times N_{G}}$ is the sensing matrix. In (\ref{CS}), the CE problem is reformulated as a sparse recovery task of the paths’ support set and complex gains. Hence, several classical CS techniques, such as SBL, can be used to estimate the sparse vector $\bm{\bar{\beta}}_{k}$ \cite{tipping2001sparse}. However, in the CS framework \cite{bajwa2010compressed}, stable channel recovery accuracy is guaranteed by minimizing the column coherence $\mu$ of the polar-domain transform matrix $\mathbf{B}$, which is defined as \cite{cui2022channel,bajwa2010compressed}
\begin{equation}
 \mu \overset{\Delta}{=} \text{max}_{p \neq q}f \left ( \vartheta _{p}, \vartheta _{q}, \gamma _{p}, \gamma _{q},r _{p}, r _{q} \right ), 
 \label{15}
\end{equation}
where $f \! \left ( \! \vartheta _{p}, \! \vartheta _{q}, \! \gamma _{p}, \! \gamma _{q}, \! r _{p}, \! r _{q} \right ) \! \overset{\Delta}{=} \! \left | \mathbf{b}\left ( \vartheta _{p},\gamma _{p},r _{p} \right )^{H} \! \mathbf{b}\left ( \vartheta _{q},\gamma _{q},r _{q} \right ) \right |$ characterizes the column coherence between two near-field array response vectors in the polar-domain transform matrix. \par
Crucially, the effectiveness of a low column coherence design is predicated on the mmWave near-field channel being sufficiently sparse in the polar domain, such that a polar-domain transform matrix $\mathbf{B}$, with finely sampled near-field array response vectors, yields well-separated sampling points in the space. Although increasing the size of $\mathbf{B}$ and reducing its coherence both tend to improve estimation accuracy, finer angular and distance sampling usually increases column coherence of the sensing matrix, revealing an intrinsic trade-off between dictionary size and recoverability. To mitigate this, we propose effective angular and distance sampling methods to reduce $\mu$ of the polar-domain transform matrix.
\subsection{Column Coherence of the Polar-Domain Transform Matrix}
\begin{figure}[!t]
\centering 
\includegraphics[width=8.5cm]{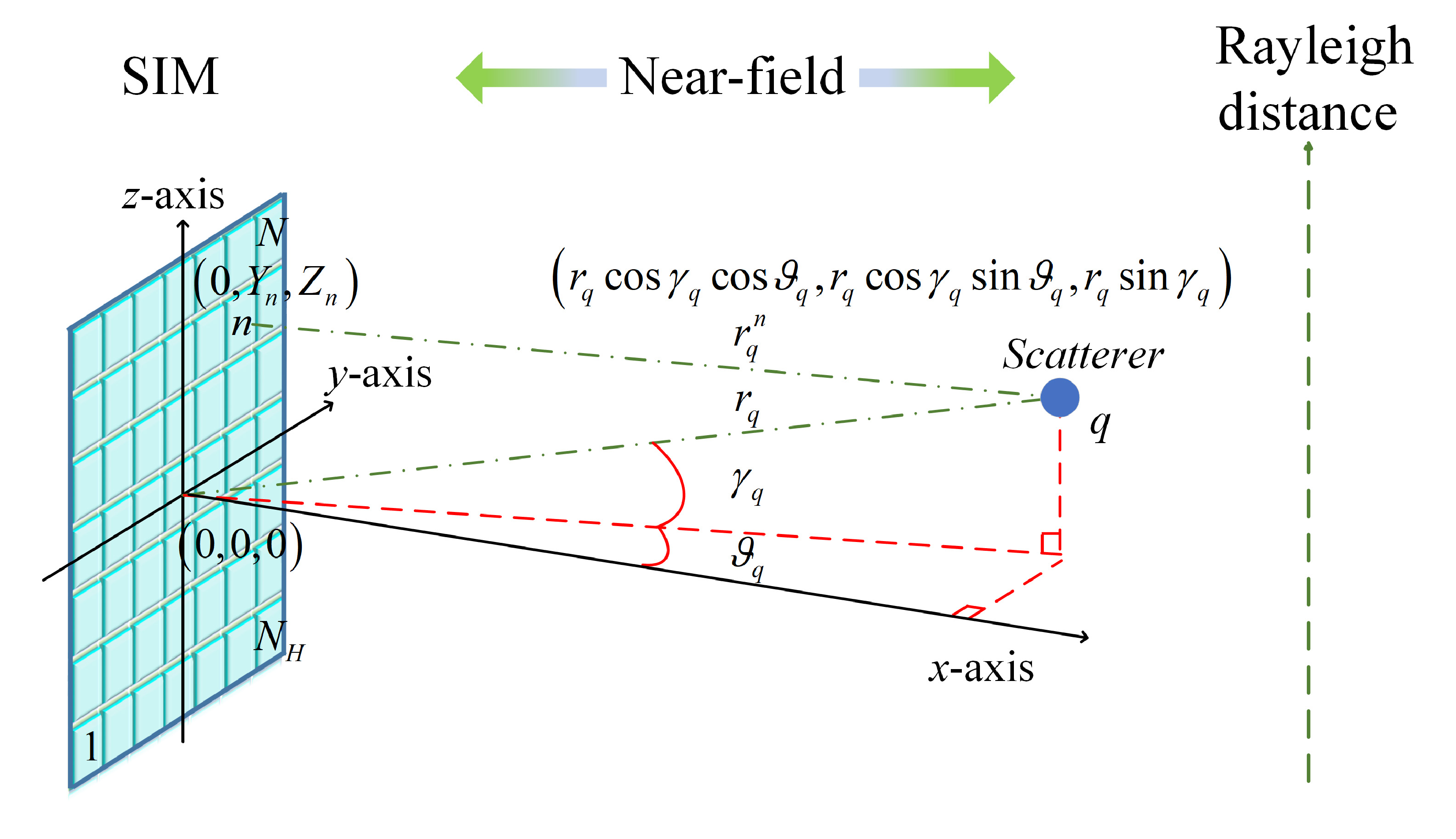} 
\caption{The near-field channel model with the $q$-th path.}
\label{Channel}
\end{figure}
As shown in Fig. \ref{Channel}, by setting the center of the last SIM layer as the origin, the coordinates of the $n$-th meta-atom relative to this origin are given by $\left ( 0, Y_n, Z_n \right )$. Moreover, $Y_n$ and $Z_n$ denote the horizontal and vertical distances of the $n$-th meta-atom from the origin, respectively, which can be calculated as 
\begin{align}
 Y_{n} &= \left ( H\left( n \right)-\frac{N_{H}+1}{2} \right )\Delta,\\
 Z_{n} &= \left ( V\left( n \right)-\frac{N_{V}+1}{2} \right )\Delta,
\end{align}
where $H\left( n \right) \!= \! \text{mod}\left ( n-1, N_{H}\right ) \! + \! 1$ and $V\left( n \right)\! = \! \left \lceil n/N_{H} \right \rceil$
denote the horizontal and vertical indices of meta-atom $n$, respectively. For simplicity, we define $\nu \overset{\Delta}{=} \text{sin}\left (\gamma \right )$, and $\iota \overset{\Delta}{=}\cos \left( \gamma \right) \sin \left ( \vartheta\right ) $. Therefore, the propagation distance between the $q$-th scatterer and the $n$-th meta-atom of the last SIM layer can be given by\footnote{It is worth noting that when the propagation distance is large, the last term in (\ref{distance}) can be neglected, which yields the far-field array response vector.}
\begin{equation}
\begin{aligned}
\label{distance}
 r_{q}^{n}&=\sqrt{r_{q}^2-2r_{q}\iota_{q} Y_{n}-2r_{q}\nu_{q}Z_{n}+Y_{n}^2+Z_{n}^2 }\\
&\overset{\left(a\right)}{\approx}r_{q}-\iota_{q}Y_{n}-\nu_{q}Z_{n}+\frac{Y_{n}^2+Z_{n}^2}{2r_{q}}, 
\end{aligned}
\end{equation}
where $\overset{\left(a\right)}{\approx}$ is derived by applying the Taylor approximation $\sqrt{1+x}\approx 1+\frac{x}{2} $\cite{haghshenas2024parametric,ziomek1993three}.

Inserting (\ref{distance}) into the function $f\left( \cdot \right)$ defined in (\ref{15}), we obtain 
\begin{equation}
\label{f1}
\begin{aligned}
 &\bar{f} \left ( \nu _{p}, \nu _{q}, \iota _{p}, \iota _{q}, r _{p}, r _{q} \right ) \\& \! \approx \! \frac{1}{N}\left | \sum_{n=1}^{N}e^{j\frac{\pi \left ( Y_{n}^2+Z_{n}^2 \right ) }{\lambda }\left ( \frac{1}{r_{p}} \!- \! \frac{1}{r_{q}} \right )-j\frac{2\pi Y_{n}}{\lambda }\left ( \iota _{p}-\iota _{q} \right )-j\frac{2\pi Z_{n}}{\lambda }\left ( \nu _{p}-\nu _{q} \right )} \right |.
\end{aligned}
\end{equation}
Therefore, by adopting appropriate angular and distance sampling methods, we can reduce the column coherence between any two distinct near-field array response vectors in the polar-domain transform matrix $\mathbf{B}$, thus improving the CS-based CE performance. However, it can be observed that obtaining angular and distance sampling methods directly from (\ref{f1}) is challenging \cite{cui2022channel}. To facilitate the subsequent analysis, next we will decouple the exponent term in (\ref{f1}) into two parts and obtain the individual sampling methods to minimize the column coherence of matrix $\mathbf{B}$.
\subsection{Proposed Sampling Method}
In this subsection, we present the angular and distance sampling methods to generate $N_G$ sampling points in the polar-domain transform matrix.
\begin{lemma}[Angular Sampling]
\label{L1}
 For a given propagation distance, sampling the azimuth and elevation AoAs 
 \begin{subequations}
 \begin{align}
 \label{nu}
 &\nu _{i}=\frac{i\lambda -N_{V}\Delta}{N_{V}\Delta},i=1,2,\dots,\left \lfloor \frac{2N_{V}\Delta}{\lambda } \right \rfloor , \\
 \label{Xi}
 &\iota _{i}=\frac{i\lambda -N_{H}\Delta}{N_{H}\Delta},i=1,2,\dots,\left \lfloor \frac{2N_{H}\Delta}{\lambda } \right \rfloor , \\
 \label{all}
 &\iota_{i}^2+\nu_{i}^2\le 1,
 \end{align}
 \end{subequations}
 yields that the column coherence asymptotically approaches zero. \par
 \textit{Proof:}
 Please refer to Appendix \ref{sam_angular}.
\end{lemma}\par
It is important to note that the angular sampling method for ULAs is a special case of Lemma 1. When the spacing is configured as $\Delta=\lambda/2$, the difference between any two elevation AoAs is $\left | \sin\left(\gamma_p\right)-\sin\left(\gamma_q\right) \right |= \frac{2i}{N_V}$, for $i=1,2,\dots,N_V $, which aligns with the existing angular sampling method designed for ULAs \cite{cui2022channel}.\par
Unfortunately, the quadratic phase in (\ref{f1}) makes it difficult to derive the explicit distance sampling method. Therefore, we introduce a control parameter to obtain the distance sampling method to ensure the column coherence is below a given threshold.
\begin{lemma}[Distance Sampling]
\label{L2}
 For an arbitrary threshold $\epsilon$, one can find a control parameter $\kappa_{\epsilon}$ to satisfy
 \begin{equation}
 \label{G_kappa}
 \stackrel\frown{G} \left ( \kappa_{\epsilon} \right )=\left | G\left ( \varpi_{H}\right ) \right |\times \left | G\left ( \varpi_{V}\right ) \right |\le \epsilon,
 \end{equation}
 where $\stackrel\frown{G} \left ( \kappa_{\epsilon} \right )$ is the theoretical column coherence of the polar-domain transform matrix corresponding to $ \kappa_{\epsilon}$, $\varpi_{H} =N_{H}\Delta\sqrt{\frac{ \kappa_{\epsilon} }{2\lambda} }$, $\varpi_{V} =N_{V}\Delta\sqrt{\frac{ \kappa_{\epsilon} }{2\lambda} }$, while $G\left ( \cdot\right )$ is given in Appendix \ref{sam_distance}. If the sampling distances satisfy
 \begin{equation}
 \label{r}
 r_{i}=\frac{1}{i\kappa _{\epsilon }}, i=1,2,\dots,
 \end{equation}
 the column coherence of any two near-field array response vectors sampled at different distances is guaranteed to be less than or equal to $\epsilon$.
The proposed design of the polar-domain transform matrix differs from those in \cite{demir2023new,cui2022channel}, since the latter requires searching for each angular pair $\left(\nu,\iota \right)$ to satisfy the distance ring $\frac{\left(1-\nu \right)^2\left( 1-\iota \right)^2}{r}=c$, where $c$ is a constant.
\par
 \textit{Proof:}
 Please refer to Appendix \ref{sam_distance}.
\end{lemma}
\begin{figure*}[!t] 
\centering
\subfloat[Sampling points.] 
{\label{Figure5}\includegraphics[width=6.25cm]{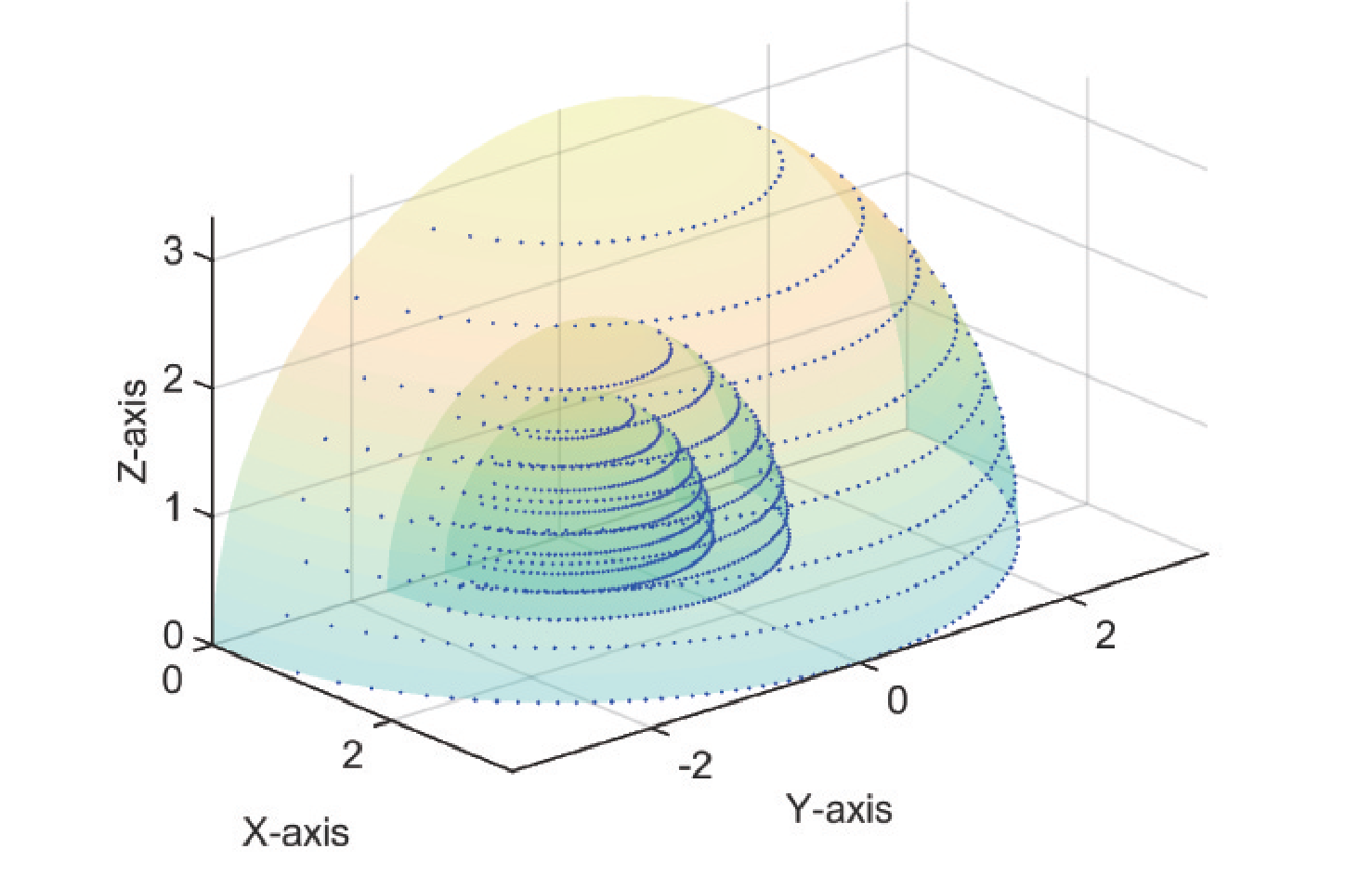}}
\subfloat[Sampling points for $\gamma=0$.] 
{\label{Figure7}\includegraphics[width=6.25cm]{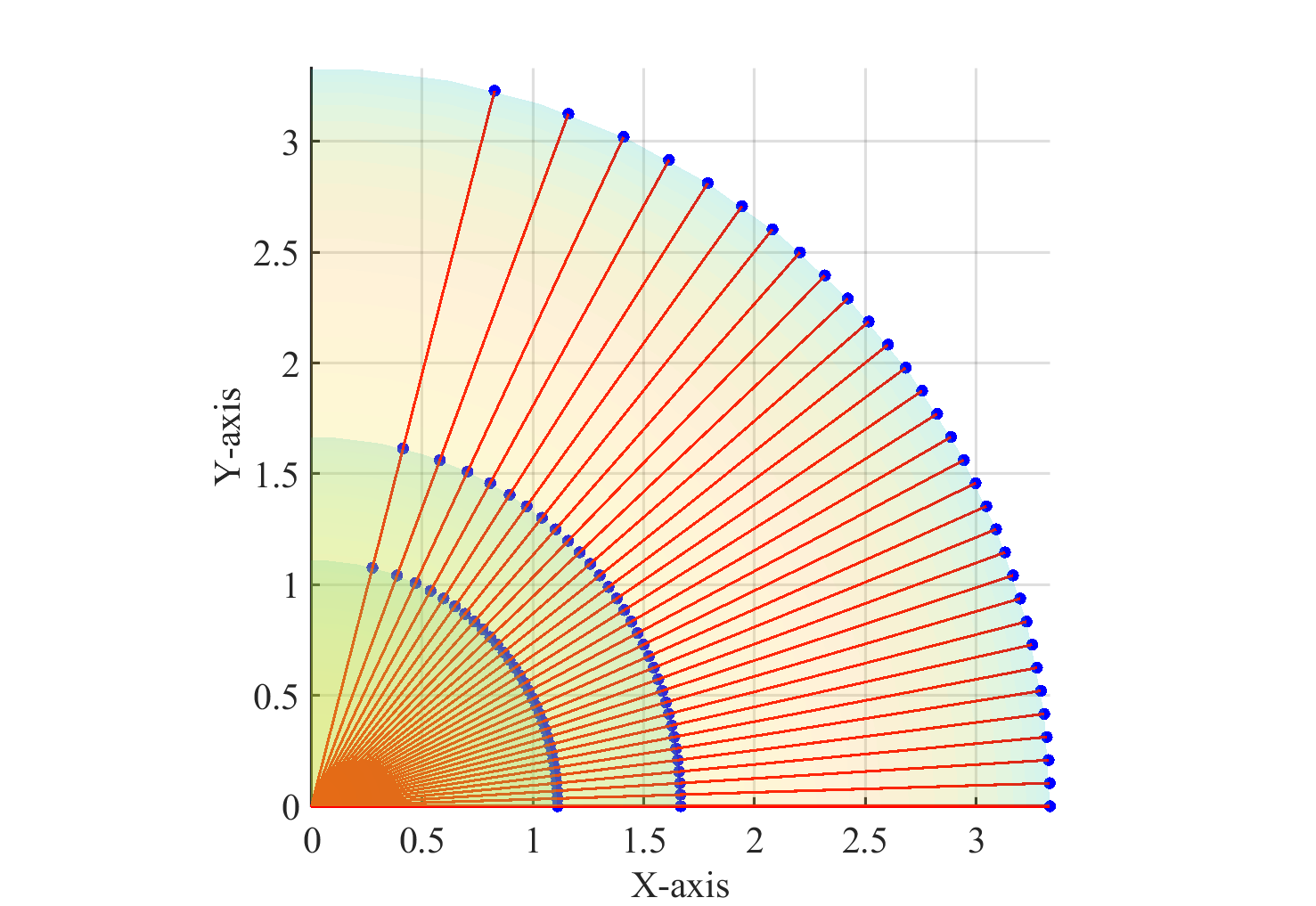}}
\subfloat[Sampling points for $\vartheta=0$.]
{\label{Figure8}\includegraphics[width=6.25cm]
{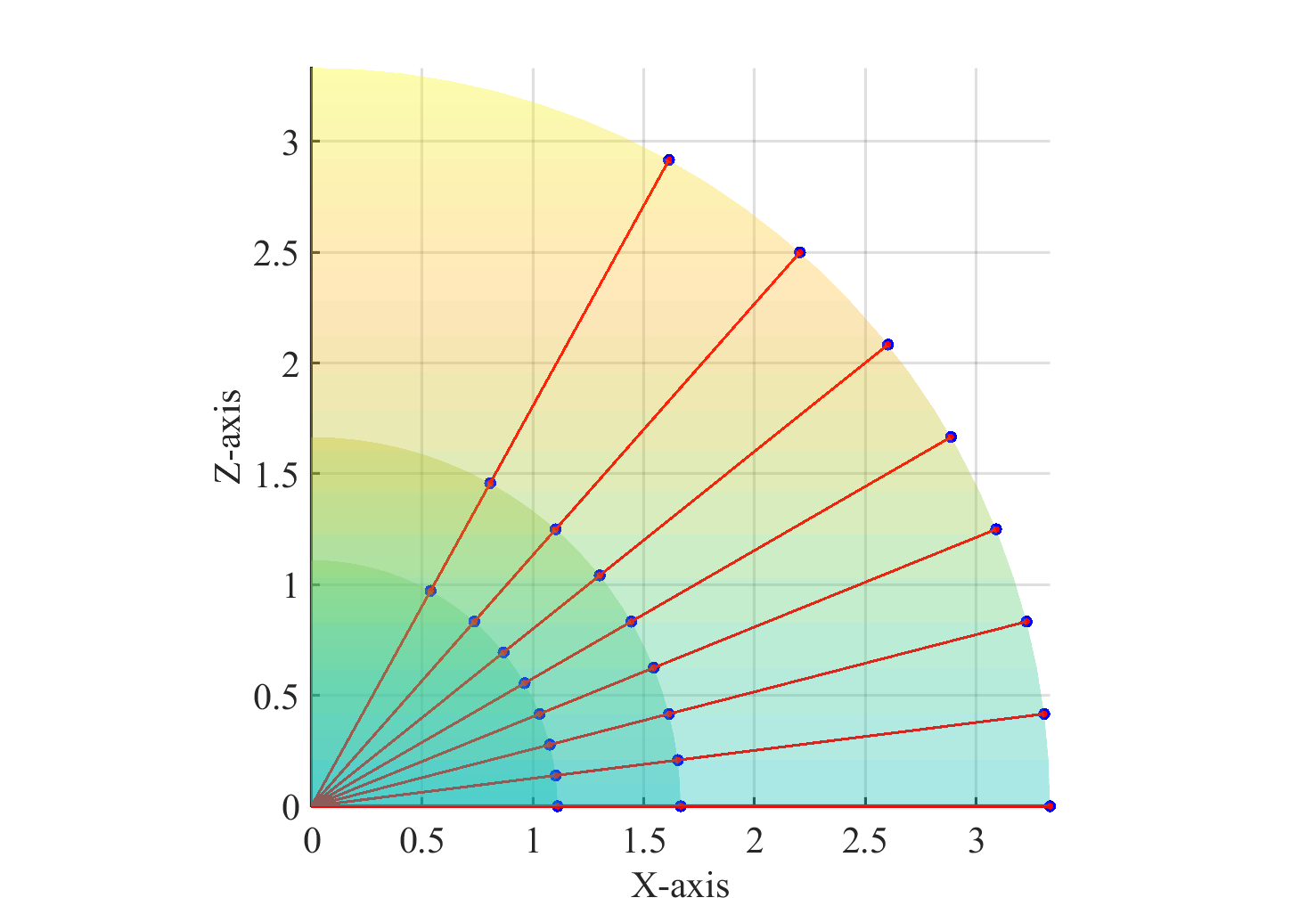}} 
\caption{Illustration of the sampling points based on the proposed polar-domain codebook for UPAs, where the blue points denote the sampling points, and the red lines denote the sampling azimuth and elevation angles, respectively.} 
\label{fig_4} 
\end{figure*}

\begin{algorithm}[!t]
 \caption{The generation of the polar-domain transform matrix $\mathbf{B}$}
 \label{transform}
 \renewcommand{\algorithmicrequire}{\textbf{Input:}}
 \renewcommand{\algorithmicensure}{\textbf{Output:}}
 \begin{algorithmic}[1]
 \REQUIRE Number of meta-atoms in each row $N_{H}$; number of meta-atoms in each column $N_{V}$; meta-atom spacing $\Delta$; wavelength $\lambda$; threshold $\epsilon$. %%input
 \ENSURE Polar-domain transform matrix $\mathbf{B}$. %%output
 \STATE Compute the Fresnel and Rayleigh distance $r_{F}$ and $r_{R}$.
 \STATE Compute the control parameter $\kappa_{\epsilon}$ from (\ref{G_kappa}).
 \STATE Initialize $u=0$; $i_r=0$; $r=1/\left( i_r\kappa_{\epsilon} \right)$.
 \WHILE{$r> r_{F}$ \& $r <r_{R}$} \label{r_F}
 \FOR{$i=1,2,\dots,\left \lfloor \frac{2N_{V}\Delta}{\lambda } \right \rfloor $}
 \STATE Select $\nu_{i}$ from (\ref{nu}) \label{nu_a}.
 \FOR{$j=1,2,\dots,\left \lfloor \frac{2N_{H}\Delta}{\lambda } \right \rfloor $}
 \STATE Select $\iota_{j}$ from (\ref{Xi}) \label{Xi_a}.
 \IF {$\nu_{i}^2+\iota_{j}^2\le 1$}
 \STATE $u \leftarrow u+1$.
 \STATE Construct $\mathbf{b}_{u}\left ( \nu_{i},\iota_{j},r \right ) $ from (\ref{b}).
 \ENDIF
 \ENDFOR
 \ENDFOR
 \STATE $i_{r} \leftarrow i_{r}+1$; $r=1/\left ( i_{r} \kappa_{\epsilon} \right ) $
 \ENDWHILE
 \STATE $\mathbf{B}=\left [ \mathbf{b}_{1}, \mathbf{b}_{2},\dots,\mathbf{b}_{u}\right ] $; $N_{G}=u$.
 \RETURN $\mathbf{B}$.
 \end{algorithmic}
\end{algorithm}
\subsection{Design of the Polar-Domain Transform Matrix}
According to the above observations, the polar-domain transform matrix's column coherence is guaranteed to be lower than the threshold $\epsilon$ when the angular pairs $\left(\nu,\iota \right)$ satisfy the uniform sampling method in (\ref{nu}), (\ref{Xi}), (\ref{all}), and the distance $r$ satisfies the non-uniform sampling method in (\ref{r}).\par 
For illustration purposes, Fig. \ref{fig_4} shows the sampling points for a $128 \times 32$ metasurface corresponding to the last layer of the SIM, with a carrier frequency of $\SI{30}{GHz}$. Additionally, the spacing between adjacent meta-atoms is set to $\Delta=\lambda/4$. Fig. \ref{Figure5} demonstrates that the sampling points are located on multiple spheres with the center at the last SIM layer and the radius determined by the non-uniform distance sampling method. Figs. \ref{Figure7} and \ref{Figure8} show that the angular sampling points are uniformly distributed on each ring.\par
After determining the polar-domain transform matrix, CE can be reformulated by recovering the propagation paths’ array response vector and complex gains. Nonetheless, the CE accuracy is limited by the resolution of the polar-domain transform matrix, which can be represented by the number of sampling points $ N_G = i_a \times i_r $, where $i_a$ is the number of angular samples on each sphere, and $i_r$ is the number of spheres along the distance dimension. In this paper, we limit the range of the distance sampling points to $ \left ( r_F, r_R \right ) $, where $r_F$ and $r_R$ denote the Fresnel and Rayleigh distances in the SIM-based mmWave near-field communication system \cite{you2023near}. The proposed polar-domain transform matrix's generation procedure is summarized in \textbf{Algorithm 1}.
\section{Proposed Near-Field CE Scheme}
\label{CS_scheme}
In this section, we propose a CE scheme within the SBL framework by leveraging the near-field channel sparsity in the polar domain. Furthermore, the LCPD-SBL algorithm is developed by replacing the inversion computation with a linear system solver to reduce the computational and spatial complexity.

\subsection{Proposed SBL-Based CE Scheme}
\begin{figure}[t]
\centering 
\includegraphics[width=8.5cm]{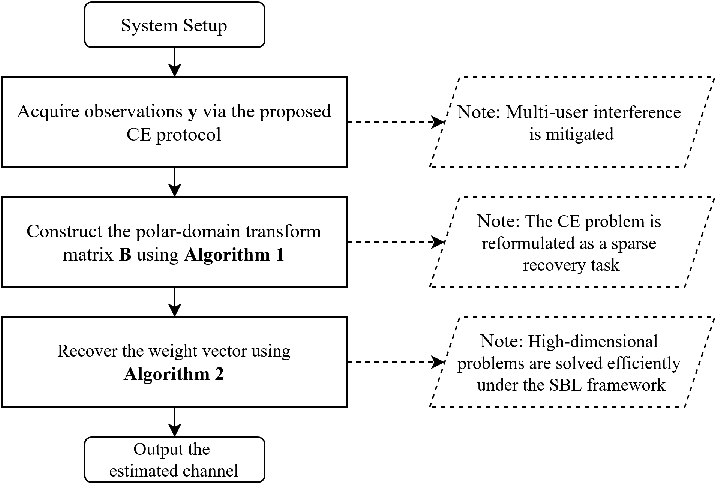} 
\caption{Flowchart of the proposed near-field CE scheme.}
\label{flowchart}
\end{figure}

As shown in Fig. \ref{flowchart}, by utilizing the proposed CE protocol in Section II-C, the CE problem for each UE is independent of each other. Without loss of generality, next we will consider an arbitrary UE. Based on the polar-domain channel representation in Section III-A, the CE problem is transformed into a sparse recovery task of the paths’ support set and complex gains. In this paper, the advanced SBL framework is invoked to achieve superior sparse signal recovery.\par
Normally, in the SBL framework, a parameterized Gaussian prior is assigned to the weight vector $\bm{\bar{\beta}}_{k}$ as follows \cite{tipping2001sparse,mishra2017sparse}
\begin{equation}
\label{SBL2}
 p\left ( \bar{\bm{\beta}}_{k}| \bm{\omega} \right )=\prod_{i=1}^{N_{G}}\left ( 2\pi\omega_{i} \right )^{-\frac{1}{2} }e^{-\frac{\bar{\beta}_{k,i}^2}{2\omega_{i}} }, 
\end{equation}
where $\bm{\omega}=\left [ \omega_{1};\omega_{2};\dots;\omega_{N_{G}} \right ] \in \mathbb{C}^{N_{G} \times 1} $, while the hyperparameter $\omega_{i}$ denotes the variance corresponding to the $i$-th element of the weight vector $\bar{\beta}_{k,i}$. It is worth noting that the Gaussian prior in (\ref{SBL2}) is imposed on the entire coefficient vector and does not require prior knowledge of the channel's sparsity level or support set. Moreover, as $\omega_{i} \to 0$, the corresponding coefficient $\bar{\beta}_{k,i} \to 0$, indicating that the hyperparameter vector $\bm{\omega}$ naturally encodes the channel sparsity. Therefore, estimating $\bar{\bm{\beta}}_{k}$ reduces to estimating the hyperparameter vector $\bm{\omega}$. \par
Generally, the expectation-maximization algorithm is utilized to estimate the hyperparameters within the SBL framework \cite{mishra2017sparse,lin2022covariance}. Specifically, the fundamental idea of the expectation-maximization algorithm involves alternating iterations of the Expectation step (\textbf{E-step}) and the Maximization step (\textbf{M-step}). The \textbf{E-step} in the $t$-th iteration treats $\bar{\bm{\beta}}_{k}$ as a hidden variable, and computes the log-likelihood function $\mathcal{L}\left (\bm{\omega}| \bm{\omega}^{\left ( t-1 \right ) } \right )$
 \begin{equation}
 \label{EM1}
 \mathcal{L}\left (\bm{\omega}| \bm{\omega}^{\left ( t-1 \right ) } \right ) =\mathbb{E}_{p\left (\bar{\bm{\beta}}_{k}|\tilde{\mathbf{y}}_{k},\bm{\omega}^{\left(t-1\right)}\right)}\left \{ \text{log}p\left (\tilde{\mathbf{y}}_{k},\bar{\bm{\beta}}_{k}|\bm{\omega}\right) \right \}, 
 \end{equation}
where $\bm{\omega}^{\left ( t-1 \right )}$ denotes the solution of the $\left(t-1 \right)$-th iteration. In the \textbf{M-step}, the $t$-th iteration's solution $\bm{\omega}^{\left ( t\right )}$ is obtained by maximizing $\mathcal{L}\left (\bm{\omega}| \bm{\omega}^{\left ( t-1 \right ) } \right )$ with respect to $\bm{\omega}$ \cite{tipping2001sparse}
 \begin{equation}
 \label{EM2}
 \begin{aligned}
 &\bm{\omega}^{\left ( t \right )}= \\& \text{arg}\ \underset{\bm{\omega}}{\text{max}} \left \{ \! \sum_{i=1}^{N_{G}}\text{log} \! \left ( 2\pi \omega_{i}^{\left ( t-1 \right ) } \! \right )^{-\frac{1}{2} } \! - \! \frac{\left | \left [ \! \bm{\mu} _{\bar{\bm{\beta}}_{k}}^{\left ( t \right ) } \right ]_{i} \right |^2 \! + \! \left [ \!\bm{\Sigma}_{\bar{\bm{\beta}}_{k}}^{\left ( t \right ) } \right ]_{i,i} }{2\omega_{i}^{\left ( t-1 \right ) }} \! \right \} ,
 \end{aligned}
 \end{equation}
where $\left [ \bm{\mu} _{\bar{\bm{\beta}}_{k}}^{\left ( t \right ) } \right ]_{i}$ and $\left [ \bm{\Sigma}_{\bar{\bm{\beta}}_{k}}^{\left ( t \right ) } \right ]_{i,i}$ is the $i$-th element of the mean vector $\bm{\mu} _{\bar{\bm{\beta}}_{k}}^{\left ( t \right ) }$ and the $\left ( i,i \right ) $-th element of the covariance matrix $\bm{\Sigma}_{\bar{\bm{\beta}}_{k}}^{\left ( t \right ) }$ corresponding to the posterior distribution over the weight vector in the $t$-th iteration, respectively, that can be obtained by \cite{tipping2001sparse}
\begin{align}
 \label{mu}
 \bm{\mu} _{\bar{\bm{\beta}}_{k}}^{\left ( t \right ) }&=\frac{\bm{\Sigma}_{\bar{\bm{\beta}}_{k}}^{\left ( t \right ) }\bm{\Pi }^{H}\mathbf{\tilde{y} }_{k}}{\tau_{p}\sigma _{n}^2}, 
 \\
 \label{sigma}
 \bm{\Sigma}_{\bar{\bm{\beta}}_{k}}^{\left ( t \right ) }&=\left ( \frac{\bm{\Pi }^{H}\bm{\Pi } }{\tau_{p}\sigma _{n}^2}+{\bm{\Omega}^{\left ( t-1 \right )}}^{-1} \right )^{-1},
\end{align}
where $\bm{\Omega}^{\left ( t-1 \right )}=\text{diag}\left( \bm{\omega}^{\left ( t-1 \right )} \right)$. Note that the maximization problem in (\ref{EM2}) can be divided into $N_G$ single-variable maximization problems involving only a single element of the hyperparameter vector. Therefore, we can obtain the optimal solution for each element individually, yielding
 \begin{equation}
 \label{EM3}
 \omega_{i}^{\left ( t \right ) }=\left | \left [ \bm{\mu} _{\bar{\bm{\beta}}_{k}}^{\left ( t \right ) } \right ]_{i} \right |^2+\left [ \bm{\Sigma}_{\bar{\bm{\beta}}_{k}}^{\left ( t \right ) } \right ]_{i,i}. 
 \end{equation}
This procedure is repeated until $\left \| \bm{\omega}^{\left ( t \right ) }-\bm{\omega}^{\left ( t-1 \right ) } \right \|_{2}\le\epsilon_{s}$ or $t\ge T$, where $\epsilon_{s}$ is the tolerable threshold and $T$ is the maximum number of iterations. Finally, the weight vector $\bar{\bm{\beta}}_{k}$, with each
of its entries denoting the corresponding complex gain associated with the $k$-th UE can be estimated as $\hat{\bar{\bm{\beta}}}_{k}=\bm{ \mu }_{\bar{\bm{\beta}}_{k}}^{\left( t \right)}$. Thus, the SIM-based near-field mmWave channel from the $k$-th UE to the last SIM layer can be obtained by
\begin{equation}
 \hat{\mathbf{h}}_{k}=\mathbf{B}\hat{\bar{\bm{\beta}}}_{k}=\mathbf{B}\bm{\mu}_{\bar{\bm{\beta}}_{k}}^{\left( t \right)}.
\end{equation}
\par
Notably, many meta-atoms are deployed on each layer of the SIM to implement effective wave-domain signal processing, thereby increasing the number of the sampled near-field array response vectors $N_{G}$. Moreover, the expectation-maximization algorithm requires inversion and storage operations for the covariance matrix $\bm{\Sigma}_{\bar{\bm{\beta}}_{k}} \in \mathbb{C}^{N_{G} \times N_{G}}$ in each iteration, leading to significant computational resources.

\subsection{Proposed LCPD-SBL Algorithm}
To overcome the limitations of the SBL framework for high-dimensional problems, we introduce a CoFEM algorithm as an alternative to the traditional expectation-maximization algorithm, and accordingly propose an LCPD-SBL algorithm to reduce the computational and spatial complexity \cite{lin2022covariance}. In each iteration, the mean vector $\bm{ \mu }_{\bar{\bm{\beta}}_{k}}$ can be derived by solving a linear system, and the diagonal elements of covariance matrix $\bm{\Sigma}_{\bar{\bm{\beta}}_{k}}$ can be obtained through a diagonal estimation rule in \cite{lin2022covariance}.\par
Specifically, after initializing the hyperparameters, in each iteration, we first rewrite (\ref{mu}) as
\begin{equation}
\mathbf{A}\bm{\mu}_{\bar{\bm{\beta}}_{k}}=\mathbf{d},
\end{equation}
where $\mathbf{A}=\frac{\bm{\Pi }^{H}\bm{\Pi } }{\tau_{p}\sigma _{n}^2}+{\mathbf{\Omega}}^{-1} \in \mathbb{C}^{N_{G} \times N_{G}}$ and $\mathbf{d}=\frac{\bm{\Pi }^{H}\mathbf{\tilde{y} }_{k}}{\tau_{p}\sigma _{n}^2} \in \mathbb{C}^{N_{G} \times 1}$, transforming the computation of $\bm{\mu}_{\bar{\bm{\beta}}_{k}}$ into solving a linear system of equations.\par 
Moreover, based on the diagonal estimation rule, we can directly calculate the required diagonal elements of the covariance matrix without the need to invert the high-dimensional matrix to obtain the entire covariance matrix $\bm{\Sigma}_{\bm{\bar{\beta}}_{k}}$ and only the required covariance matrix's diagonal elements need to be stored \cite{lin2022covariance}. Specifically, we generate $C$ $N_G$-dimensional probing vectors $\mathbf{d}_{1}, \mathbf{d}_{2}, \dots, \mathbf{d}_{C}$ following the Rademacher distribution, where each element of a probing vector is equally likely to be $-1$ or $+1$ to achieve an unbiased estimation of the diagonal elements of $\bm{\Sigma}_{\bm{\bar{\beta}}_{k}}$ \cite{lin2022covariance}. As a result, the $\left ( i,i \right ) $-th element of the covariance matrix $\bm{\Sigma}_{\bar{\bm{\beta}}_{k}}$ can be estimated as
\begin{equation}
\label{sigma2}
 \left [ \bm{\Sigma}_{\bm{\bar{\beta}}_{k}} \right ]_{i,i}=\frac{1}{C} \sum_{c=1}^{C} D_{i,c}\cdot X_{i,c}, 
\end{equation}
where $D_{i,c}$ denotes the $\left ( i,c \right ) $-th element of $\mathbf{D}=\left [ \mathbf{d}_{1} , \dots , \mathbf{d}_{C}, \mathbf{d}\right ] \in \mathbb{C}^{N_{G} \times \left( C+1 \right)}$, and $ X_{i,c}$ denotes the $\left ( i,c \right ) $-th element of $\mathbf{X}=\left [ \mathbf{x}_{1} ,\dots , \mathbf{x}_{C}, \bm{\mu}_{\bm{\bar{\beta}}_{k}}\right ] \in \mathbb{C}^{N_{G} \times \left (C+1 \right)}$, where $\mathbf{X}$ is the matrix to be updated during each iteration. The mean vector $\bm{\mu}_{\bm{\bar{\beta}}_{k}}$ required for each iteration can be obtained by solving $C+1$ separate linear systems, i.e., $\mathbf{AX}=\mathbf{D}$ \cite{lin2022covariance}. To reduce the computation complexity of the LCPD-SBL algorithm, we use a parallel conjugate gradient (PCG) algorithm, which is an extension of the preconditioned conjugate gradient method for solving multiple linear systems \cite{lin2022covariance}. The proposed LCPD-SBL algorithm's details are described in \textbf{Algorithm 2}.
\begin{algorithm}[t]
 \caption{The proposed low-complexity polar-domain SBL algorithm}
 \label{LPDSBL}
 \renewcommand{\algorithmicrequire}{\textbf{Input:}}
 \renewcommand{\algorithmicensure}{\textbf{Output:}}
 \begin{algorithmic}[1]
 \REQUIRE Received signal $\tilde{\mathbf{y}}_{k} \in \mathbb{C}^{\Psi M \times 1}$; maximum number of iterations $T$; number of probe vectors $C$; tolerable threshold $\epsilon_{s}$. %%input
 \ENSURE The estimated weight vector $\hat{\bm{\bar{\beta}}}_{k}$. %%output
 \STATE Construct the transmission matrix $\mathbf{P}$ according to (\ref{CS2}).
 \STATE Construct the proposed polar-domain transform matrix $\mathbf{B}$ by applying \textbf{Algorithm 1}.
 \STATE Compute the sensing matrix $\bm{\Pi }=\sqrt{p_{k}}\tau_{p}\mathbf{P}\mathbf{B}$.
 \STATE Initialize $\omega_{i}^{\left ( -1 \right ) }=0$, $\omega_{i}^{\left ( 0 \right ) }=1 \ \text{for} \ i=1,\dots,N_{G}$; $t=0$.
 \WHILE{$t < T$ and $\left \| \bm{\omega}^{\left ( t \right ) }-\bm{\omega}^{\left ( t-1 \right ) } \right \|_{2}>\epsilon_{s} $} 
 \STATE Update $t \leftarrow t+1$.
 \STATE Define $\mathbf{A}=\frac{\bm{\Pi }^{H}\bm{\Pi } }{\tau_{p}\sigma _{n}^2}+{\mathbf{\Omega}^{\left( t-1\right)}}^{-1}$, $\mathbf{d}=\frac{\bm{\Pi }^{H}\mathbf{\tilde{y} }_{k}}{\tau_{p}\sigma _{n}^2}$.
 \STATE Draw $\mathbf{d}_{1},\mathbf{d}_{2},\dots,\mathbf{d}_{C}\sim \text {Rademacher distribution}$.
 \STATE Construct $\mathbf{D} \!= \! \left [ \mathbf{d}_{1} ,\! \dots \!,\! \mathbf{d}_{C},\! \mathbf{d}\right ]$, $\mathbf{X}\!=\! \left [ \mathbf{x}_{1} ,\! \dots ,\! \mathbf{x}_{C},\! \bm{\mu}_{\bm{\bar{\beta}}_{k}}^{\left( t \right)}\right ]$.
 \STATE Solve $\mathbf{AX}=\mathbf{D}$ by applying the PCG algorithm.
 \STATE Update $\left [ \bm{\Sigma}_{\bm{\bar{\beta}}_{k}}^{\left( t \right)} \right ]_{i,i}$ according to (\ref{sigma2}).
 \STATE Update $\omega_{i}^{\left ( t \right ) }$ according to (\ref{EM3}).
 \ENDWHILE
 \RETURN $\hat{\bm{\bar{\beta}}}_{k}=\bm{\mu}_{\bar{\bm{\beta}}_{k}}^{\left( t \right)}$.
 \end{algorithmic}
\end{algorithm}
\subsection{Complexity Comparisons}
Compared to the iterative expectation-maximization algorithm requiring $O\left(N_{G}^3\right)$-time and $O\left(N_{G}^2\right)$-space in each iteration, the proposed LCPD-SBL algorithm's complexity is reduced to $O\left( UC\tau_{D}\right)$-time and $O\left(N_{G}\right)$-space, where $U$ denotes the maximum number of iterations of the PCG algorithm, and $\tau_{D}$ denotes the time required for matrix-vector multiplication \cite{lin2022covariance}. If the condition number $\lambda_{\max}\left( \mathbf{A} \right)/\lambda_{\min}\left( \mathbf{A} \right)$ is small, we have $U \ll N_{G}$, where $\lambda_{\max}\left( \mathbf{A} \right)$ and $\lambda_{\min}\left( \mathbf{A} \right)$ denote the largest and smallest eigenvalues of $\mathbf{A}$, respectively \cite{hestenes1952methods}. However, in the iteration of \textbf{Algorithm 2}, most of the hyperparameters are equal to $0$, which would increase the condition number. To address this issue, we introduce a Jacobi preconditioner matrix $\mathbf{M}=\text{diag}\left ( \left[\mathbf{A}\right]_{1,1},\left[\mathbf{A}\right]_{2,2},\dots ,\left[\mathbf{A}\right]_{N_{G},N_{G}} \right ) $ to transform the problem of solving for $\mathbf{X}$ in $\mathbf{AX}=\mathbf{D}$ into the equivalent problem of solving for $\mathbf{X'}$ in $\mathbf{A'X'}=\mathbf{D'}$, where $\mathbf{A'}=\mathbf{M}^{-1/2}\mathbf{A}\mathbf{M}^{-1/2}$, $\mathbf{X}'=\mathbf{M}^{1/2}\mathbf{X}$, and $\mathbf{D'}=\mathbf{M}^{-1/2}\mathbf{D}$ \cite{shewchuk1994introduction,lin2022covariance}. Moreover, it has been confirmed that a small number of detection vectors $C$ is sufficient in practice \cite{lin2022covariance}. In our simulations, we set $C = 8$.

\section{Simulation Results}
\label{simulation}
In this section, we present simulation results to assess the performance of the proposed near-field CE scheme.
\subsection{Simulation Setup}
\begin{figure}[!t]
\centering 
\subfloat[{Column coherence against $\kappa$.}] 
{\label{column}\includegraphics[width=8cm]{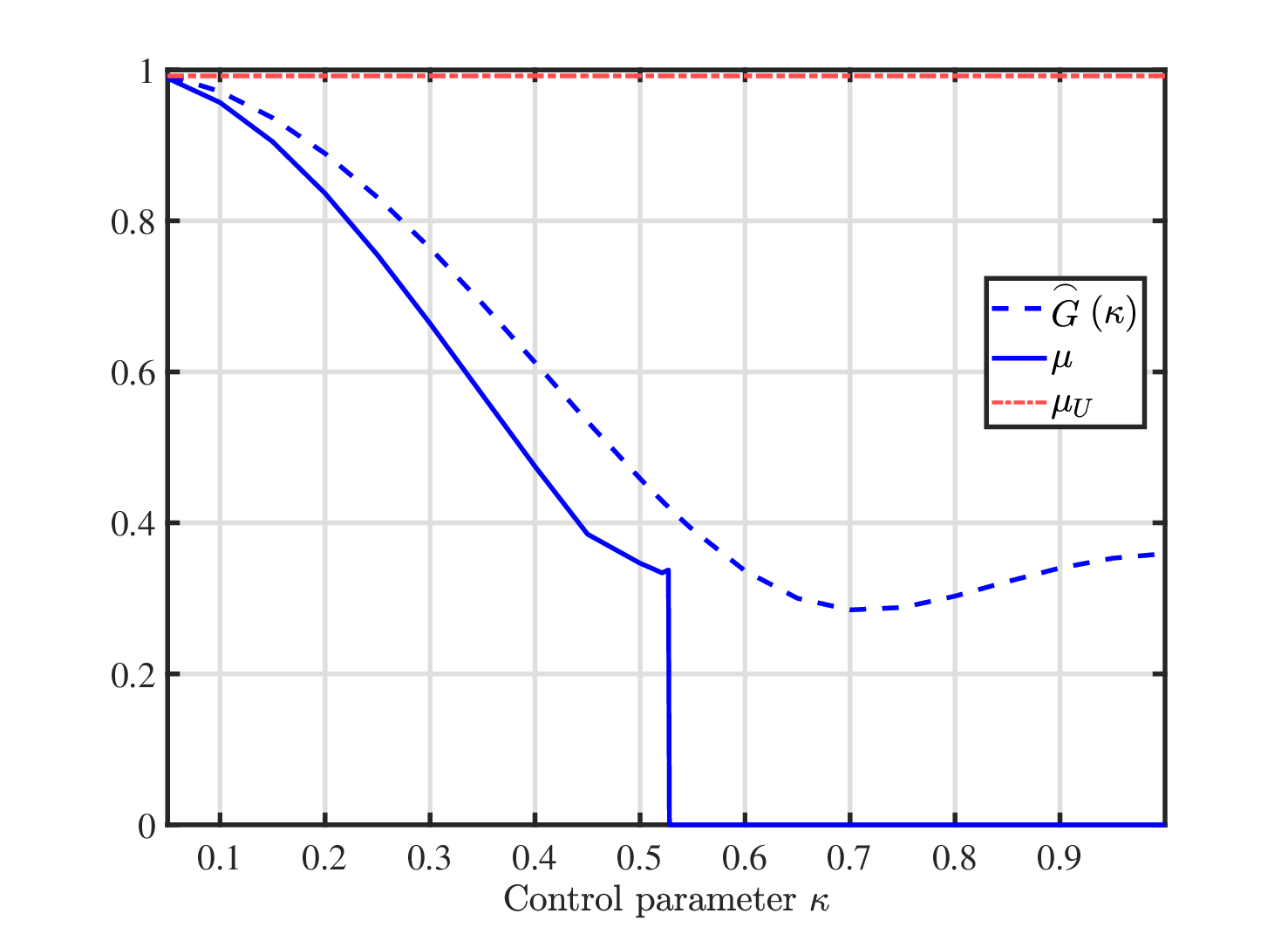}}\\ 
\subfloat[The number of samples along the distance dimension $i_{r}$ against $\epsilon$.] 
{\label{i_r}\includegraphics[width=8cm]{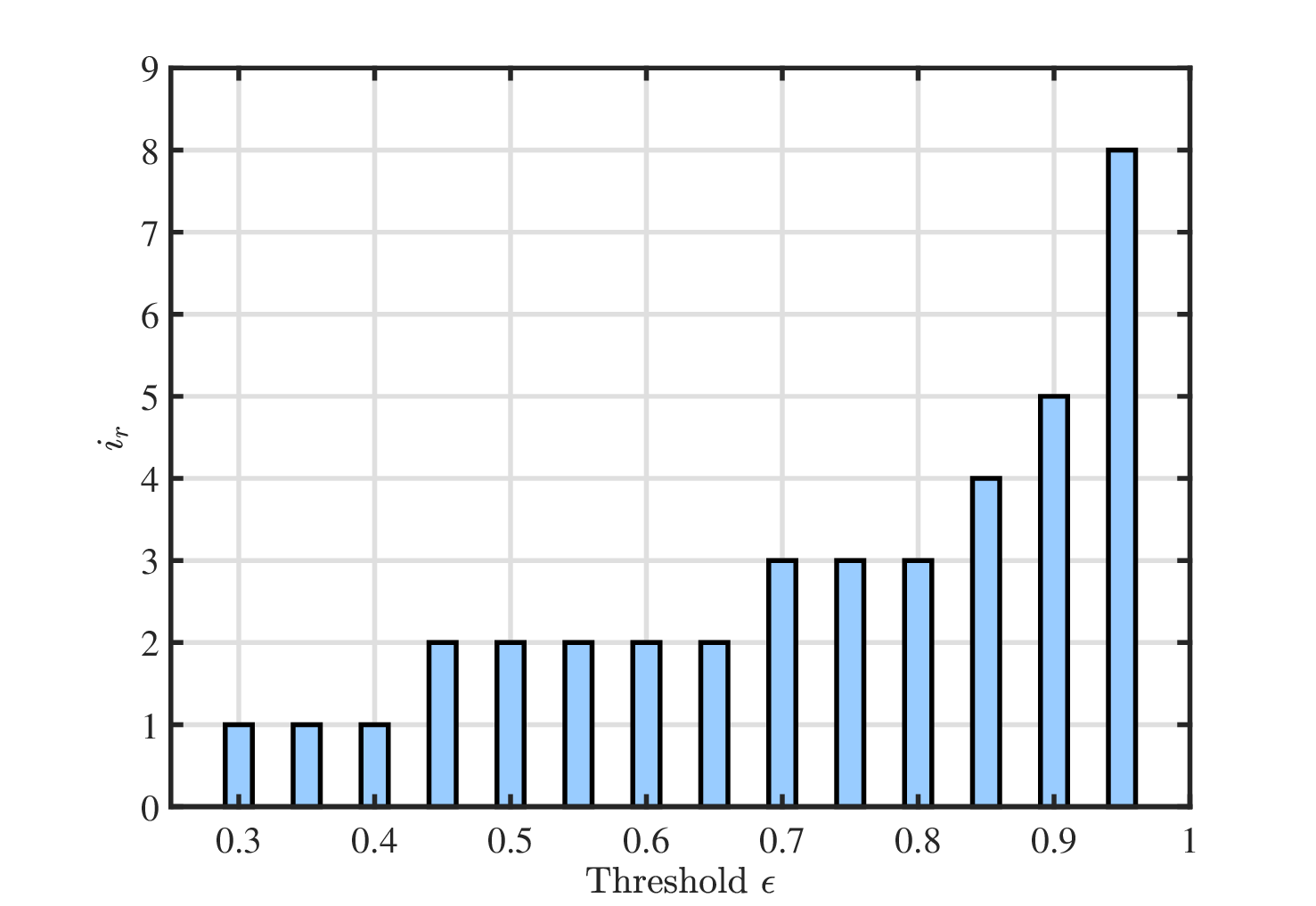}}
\caption{Verification of our distance sampling method, where we consider a $128 \times 32$ UPA working at $\SI{30}{GHz}$.}
\label{trade_off}
\end{figure}
We consider an SIM-based mmWave near-field communication system. The BS is configured with $M = 4$ antennas, and each is connected to a dedicated RF chain. The spacing between adjacent antennas is set to $\lambda/2$. Moreover, the SIM comprises four layers, while each metasurface layer is modeled as a UPA with $N_{H}=128$ meta-atoms per row, $N_{V}=32$ meta-atoms per column. Each meta-atom has dimensions of $d_{x}=d_{y}=\lambda/8$. The distance between adjacent meta-atoms is set to $\Delta=\lambda/4$. The horizontal distance between the BS and the first SIM layer and the spacing between each metasurface layer are identical and set to $d_{t}=d_{\text{Layer}}=5\lambda/L$. We assume that the phase shift of the transmission coefficient in each block satisfies $\theta _{n}^{l} \sim \mathcal{U}\left [ 0,2\pi \right )$ for $\forall n \in \mathcal{N},\forall l \in \mathcal{L}$. For the mmWave channel, the complex gain $\beta_{q,k}$ is modeled by $\mathcal{CN}\left ( 0,1/\left ( 4\pi r_{q,k} /\lambda \right )^2 \right )$ \cite{li2017millimeter}. The azimuth and elevation AoAs are both distributed as $\mathcal{U}\left ( -\frac{\pi}{3}, \frac{\pi}{3} \right )$. The number of propagation paths is $Q=6$.\par 
Additionally, the carrier frequency is $\SI{30}{GHz}$, corresponding to the wavelength of $\lambda=\SI{0.01}{m}$. Then, the array aperture is $D=\sqrt{N_{H}^2+N_{V}^2}\Delta\approx \SI{0.33}{m}$. The Fresnel distance and the Rayleigh distance are obtained as $r_{F}=\frac{1}{2} \sqrt{D^3/\lambda }\approx \SI{0.95}{m}$ and $r_{R}=2D^2/\lambda \approx \SI{21.78}{m}$. Thus, the radiating near-field region in the SIM-based mmWave communication system is $\left ( \SI{0.95}{m},\SI{21.78}{m} \right )$.  The signal-to-noise ratio (SNR) is defined as $10\log_{10}\left (  \mathbb{E}\left \{ \left \|\tilde{\mathbf{y}}-\tilde{\mathbf{n}}  \right \|^2_2 \right \}  / \mathbb{E}\left \{ \left \|\tilde{\mathbf{n}}  \right \|^2_2\ \right \}  \right) $. Furthermore, the performance is evaluated via NMSE, yielding
\begin{equation}
 \text{NMSE}= \mathbb{E}\left \{ \left \| \mathbf{h}-\mathbf{\hat{\mathbf{h}}} \right \|^{2}_{2} /\left \| \mathbf{h} \right \|^{2}_{2} \right \} ,
\end{equation} 
where $\mathbf{\hat{\mathbf{h}}}$ denotes the estimate of the true channel $\mathbf{h}$.

First, in Fig. \ref{column}, we show the practical and theoretical column coherence of the polar-domain transform matrix corresponding to the control parameter $ \kappa $. As the control parameter $\kappa$ increases, the column coherence of polar-domain transform matrix $\mu$ consistently remains below the theoretical column coherence $\stackrel\frown{G} \left ( \kappa \right )$. Additionally, we also show the column coherence under uniform distance sampling $\mu_U$ for comparison. It can be observed that the proposed non-uniform distance sampling method significantly reduces the column coherence of the polar-domain transform matrix compared to the uniform sampling approach. This demonstrates the effectiveness of our angular and distance sampling methods in controlling column coherence. Additionally, as shown in Fig. \ref{i_r}, a low threshold $\epsilon$ yields a high control parameter $\kappa$, which leads to a reduction in $i_r$ that can be selected within the range of $ \left ( r_F, r_R \right ) $. When $ \epsilon $ decreases beyond a certain threshold, the value of $ i_r $ drops to one, and the column coherence of the matrix $ \mathbf{B} $ approaches zero. In this case, all sampling points collapse onto the angular dimension, which substantially reduces the granularity of the polar-domain transform matrix $ \mathbf{B} $ along the distance dimension. This implication not only diminishes the representational richness of the transform matrix but also sharply decreases the total number of sampling points $ N_G $. Such a reduction in $ N_G $ significantly weakens the ability of $ \mathbf{B} $ to capture the near-field array manifold, thereby degrading the channel estimation accuracy. Therefore, the threshold $ \epsilon $ must be carefully chosen to strike a proper balance between implementation complexity and channel estimation performance.

\begin{figure}[!t]
\centering 
\includegraphics[width=8cm]{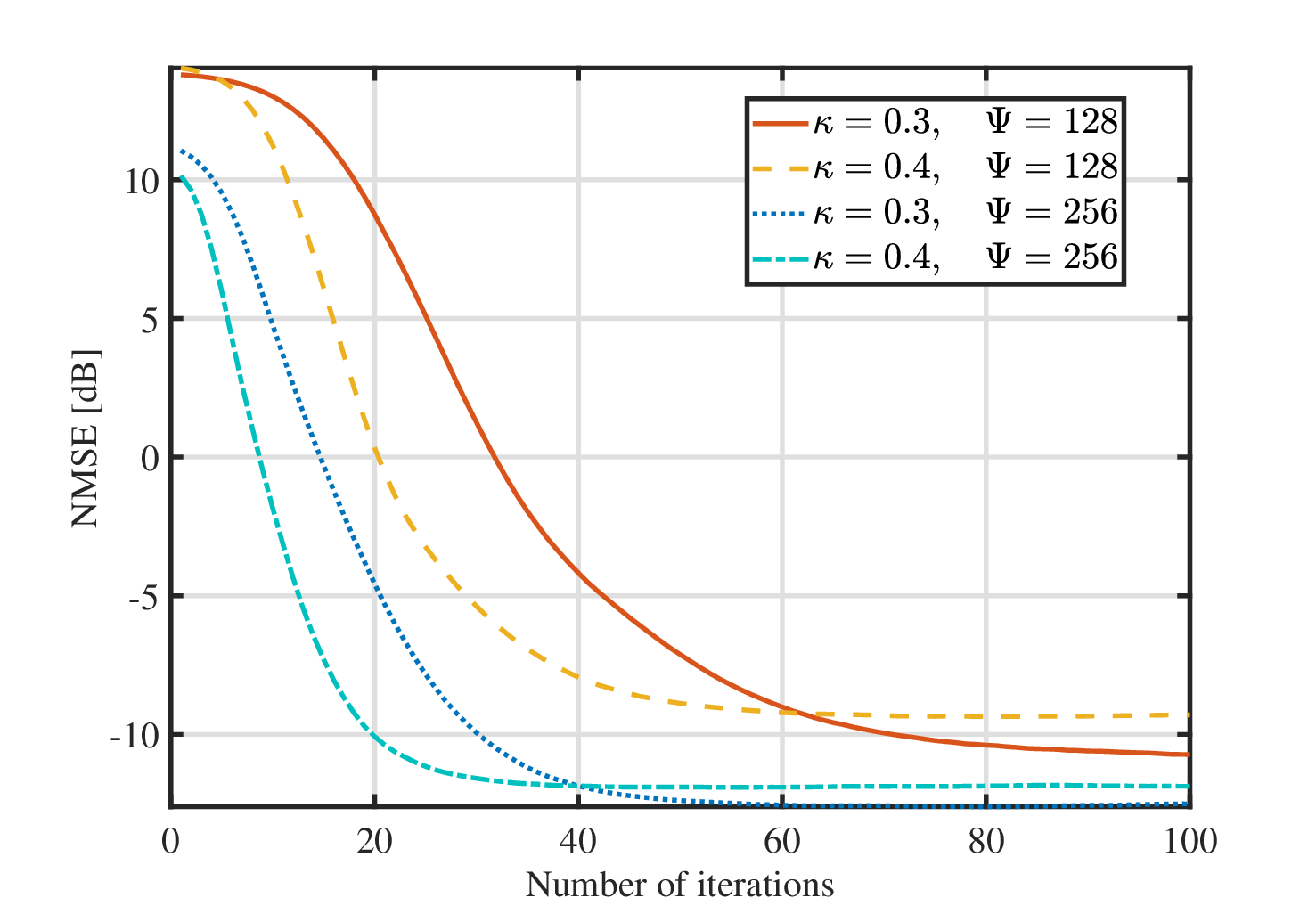} 
\caption{NMSE against the number of iterations.}
\label{convergence}
\end{figure}
\begin{figure}[!t]
\centering 
\includegraphics[width=8cm]{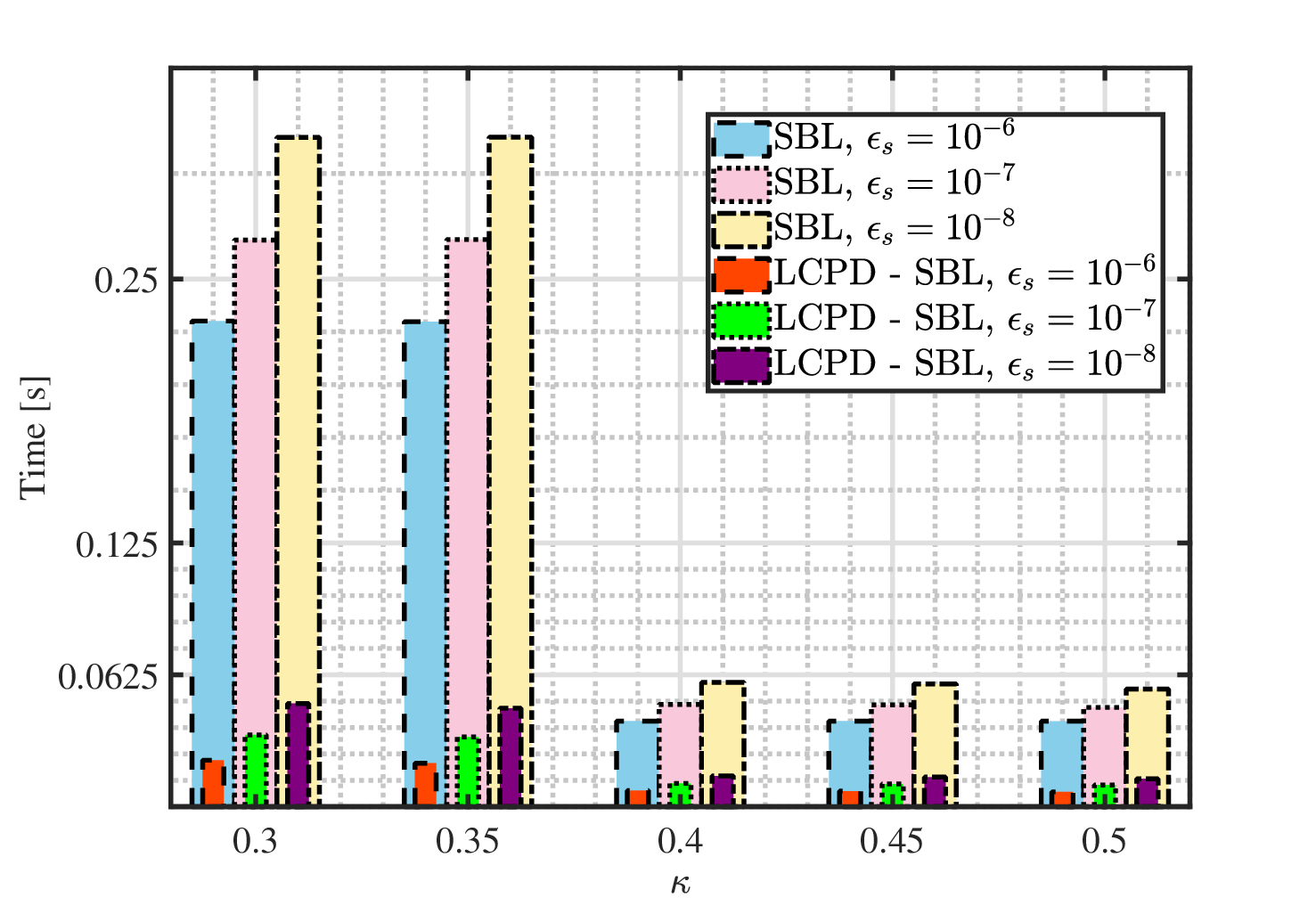} 
\caption{Computation time with respect to $\kappa$ under different tolerable thresholds.}
\label{time}
\end{figure}
\begin{figure}[!t] 
\centering
\subfloat[NMSE performance comparisons of different schemes against SNR, where multiple near-field UEs (scatterers) are located in the range of $\left ( r_{F},\SI{10}{m} \right )$.] {\label{near}\includegraphics[width=8cm]{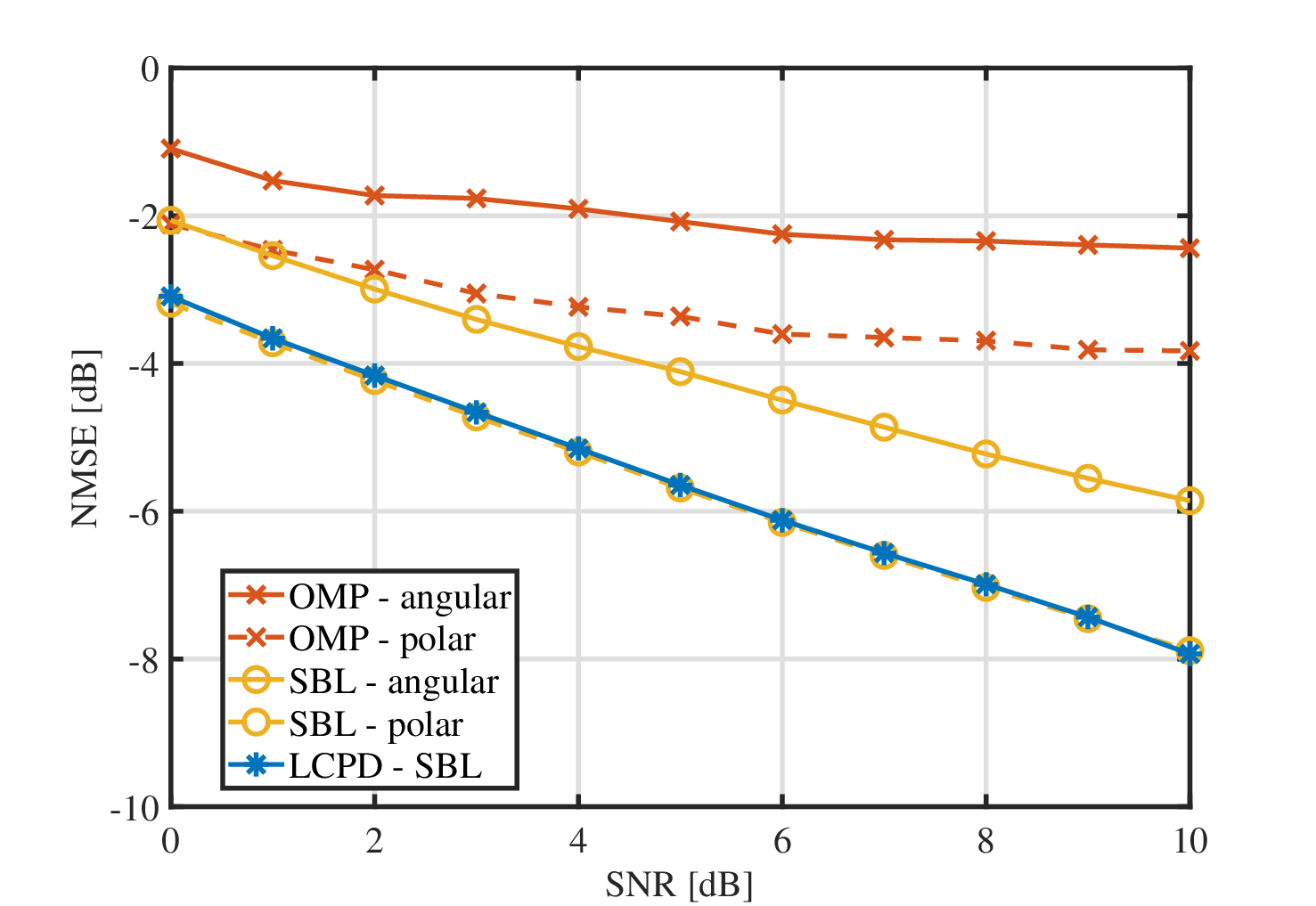}} \\ 
\subfloat[NMSE performance comparisons of different schemes against SNR, where multiple far-field UEs (scatterers) are located in the range of $\left ( \SI{80}{m},\SI{100}{m} \right )$.] 
{\label{far}\includegraphics[width=8cm]{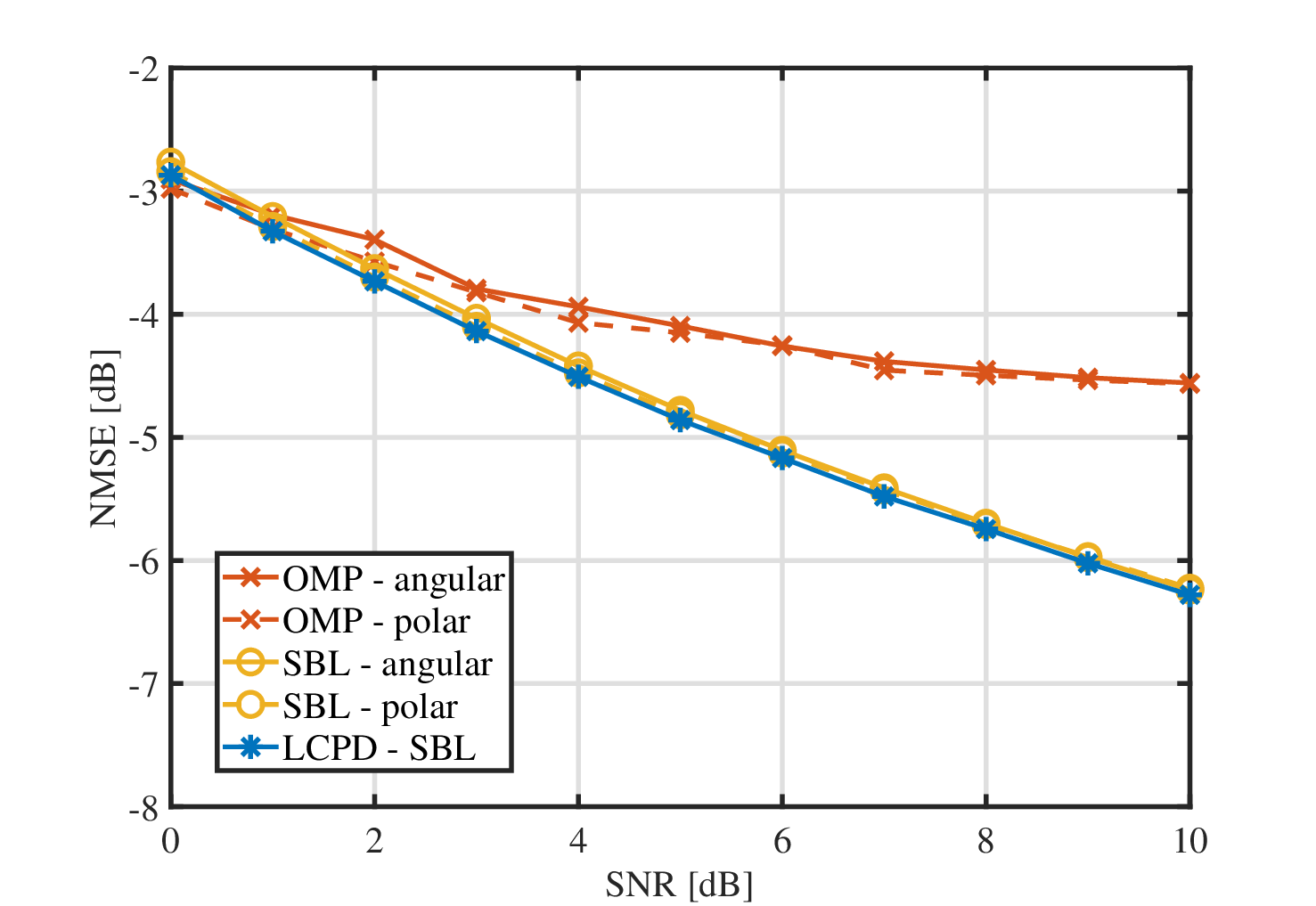}} 
\caption{NMSE performance comparison of different distances.}
\label{near_far}
\end{figure}
Then, we verify the proposed LCPD-SBL algorithm's convergence behavior in Fig. \ref{convergence}. Note that the NMSE decreases monotonically as the iterations of the LCPD-SBL algorithm proceed, which fully demonstrates that the proposed algorithm can continuously enhance the estimation accuracy during the iterations. Moreover, the proposed algorithm's convergence is guaranteed under different setups. We observe that the required number of iterations for the proposed algorithm to converge decreases as the control parameter $\kappa$ and the number of pilot blocks $\Psi$ increase.

Additionally, to verify the effectiveness of the proposed LCPD-SBL algorithm in reducing the computational complexity, Fig. \ref{time} presents the average computation time required for estimating the channel corresponding to each meta-atom, where we consider the existing SBL algorithm \cite{tipping2001sparse} and the LCPD-SBL algorithm under different setups. As the control parameter $\kappa$ increases, the computation time needed by both the SBL algorithm and the proposed LCPD-SBL algorithm decreases. This is because the increase in $\kappa$ will result in a reduction in the number of samples along the distance dimension $i_r$, thereby causing a reduced dimension $N_G$ of the weight vector $\bar{\bm{\beta}}_{k}$ that needs to be recovered in the SBL framework. Moreover, under different tolerable thresholds, the computation time for the LCPD-SBL algorithm to reach convergence is significantly lower than that of the SBL algorithm.

Considering the performance and complexity of the proposed LCPD-SBL algorithm, we set the control parameter $\kappa=0.4$. We set the number of pilot blocks to $\Psi=128$. Accordingly, the number of measurements is $128 \times M$ for an $N$-dimensional channel, leading to a compression ratio of $1/8$. The maximum number of iterations is $T=50$, while the tolerable threshold is $\epsilon_s=10^{-8}$ in the following simulations unless otherwise specified.\par
In Figs. \ref{near} and \ref{far}, we evaluate the NMSE performance with respect to the SNR for UEs (scatterers) in near-field and far-field scenarios, respectively. We compare the proposed LCPD-SBL algorithm with the existing OMP \cite{cui2022channel} and SBL \cite{tipping2001sparse} algorithms based on both the angular-domain transform matrix and the proposed polar-domain transform matrix. As shown in Fig. \ref{near}, we validate the effectiveness of the proposed CE scheme in the SIM-assisted mmWave communication system. When the UEs (scatterers) are uniformly distributed within the near-field region at a distance of ($r_{F}$, $\SI{10}{m}$), all CS-based algorithms, relying on the proposed polar-domain transform matrix, outperform their counterparts relying on the existing angular-domain transform matrix. This is because converting the near-field channel to the angular-domain representation introduces a severe energy spread effect, making the channel sparsity in the angular-domain not achievable \cite{cui2022channel}. In contrast, the proposed polar-domain transform matrix effectively enhances the channel sparsity by representing the near-field channel in the polar-domain.

\begin{figure}[!t]
\centering 
\includegraphics[width=8cm]{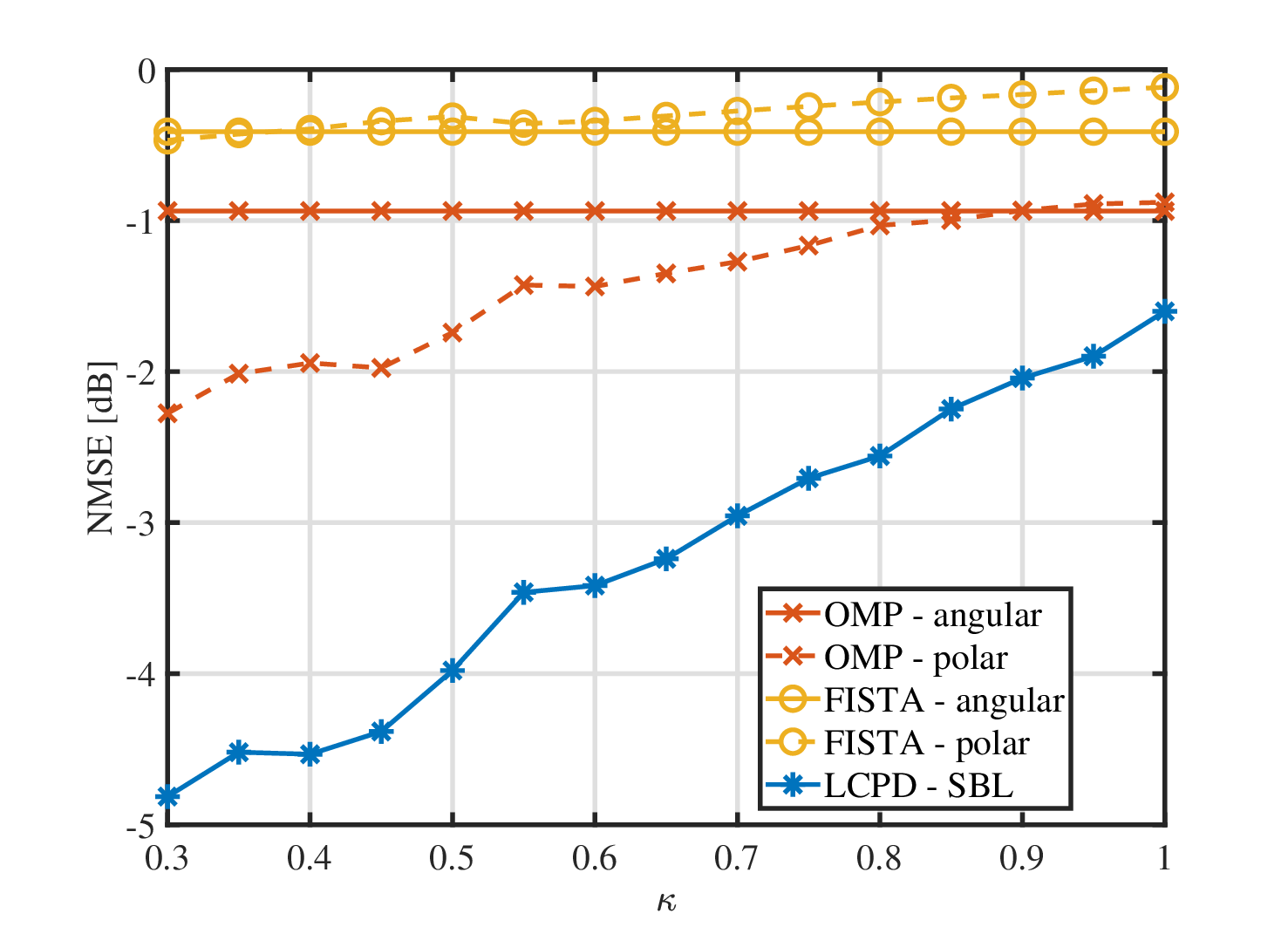} 
\caption{NMSE of different schemes against $\kappa$.}
\label{kappa_change}
\end{figure}
\begin{figure}[!t]
\centering 
\includegraphics[width=8cm]{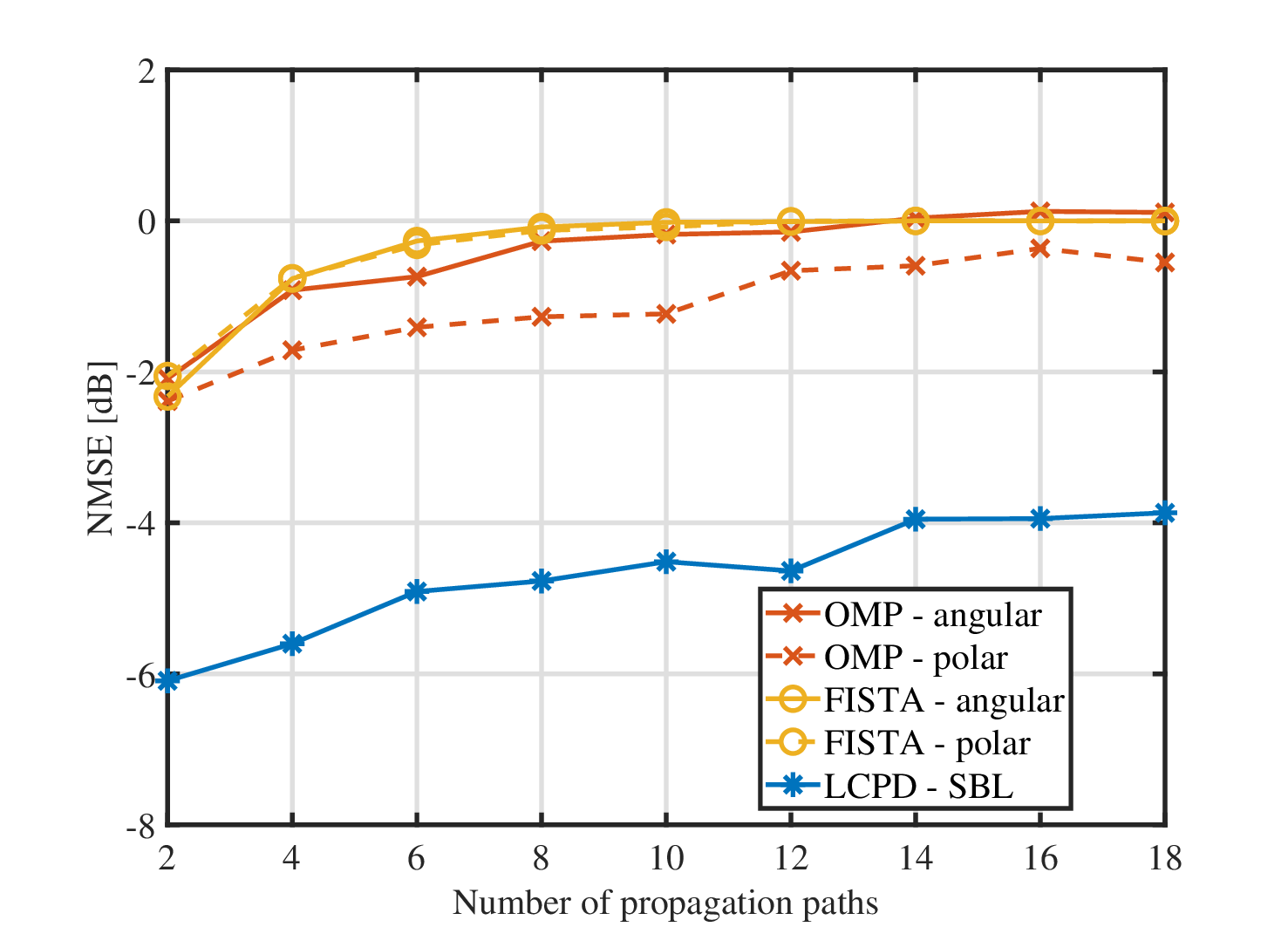} 
\caption{NMSE of different schemes against $Q$.}
\label{path_change}
\end{figure}
In Fig. \ref{far}, when the UEs (scatterers) are uniformly distributed within the far-field region at a propagation distance of ($\SI{80}{m}$, $\SI{100}{m}$), the CS-based algorithms relying on the proposed polar-domain transform matrix achieve performance similar to their counterparts relying on the existing angular-domain transform matrix. This is because, as the distance increases, the accuracy advantage of the near-field channel modeling diminishes. Nonetheless, in both near-field and far-field regions, the proposed LCPD-SBL algorithm performs similarly to the SBL algorithm based on the polar-domain transform matrix.

\begin{figure}[!t]
\centering 
\includegraphics[width=8cm]{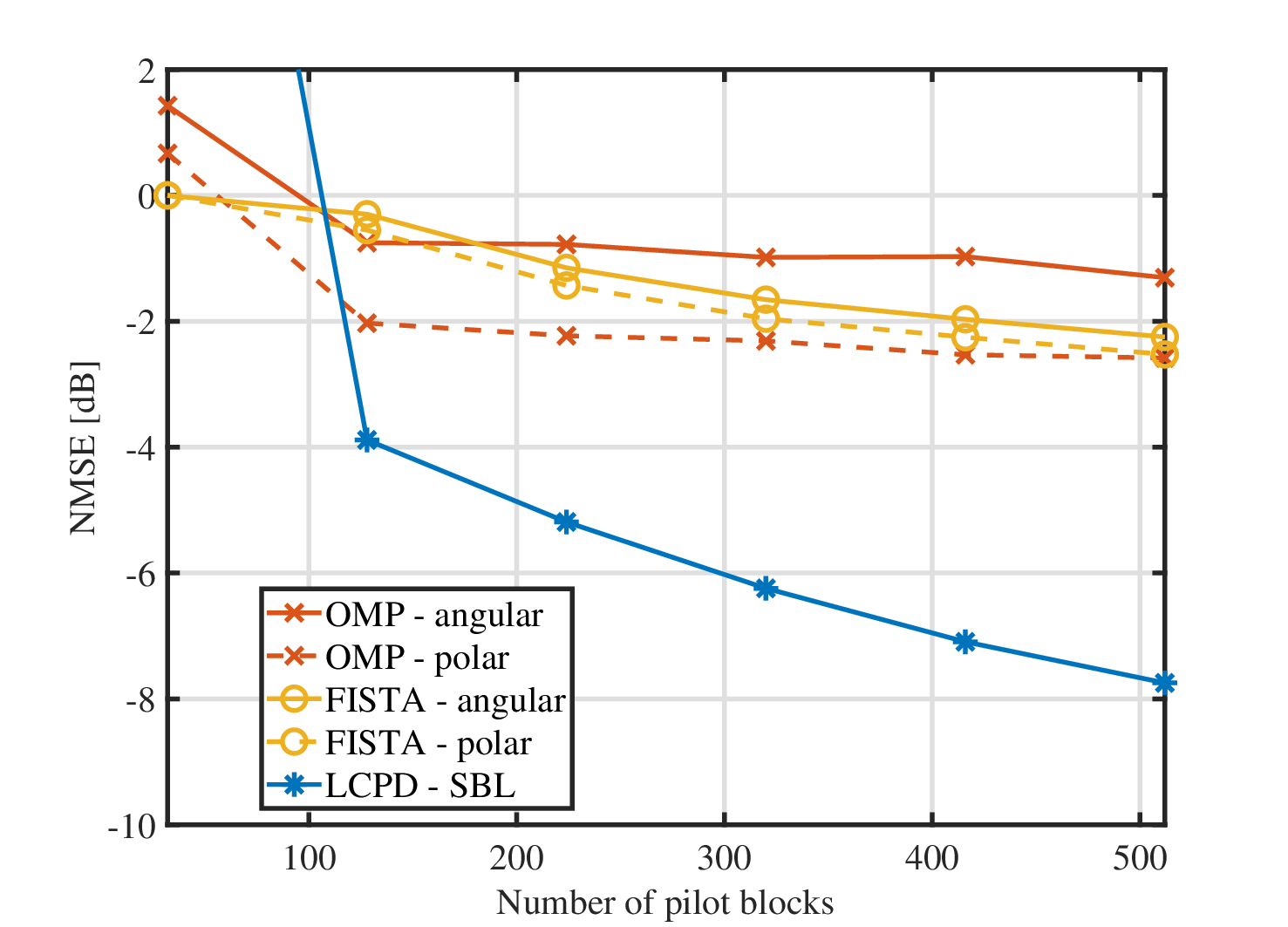} 
\caption{NMSE of different schemes against $\Psi$.}
\label{blocks_change}
\end{figure}
\begin{figure}[!t]
\centering 
\includegraphics[width=8cm]{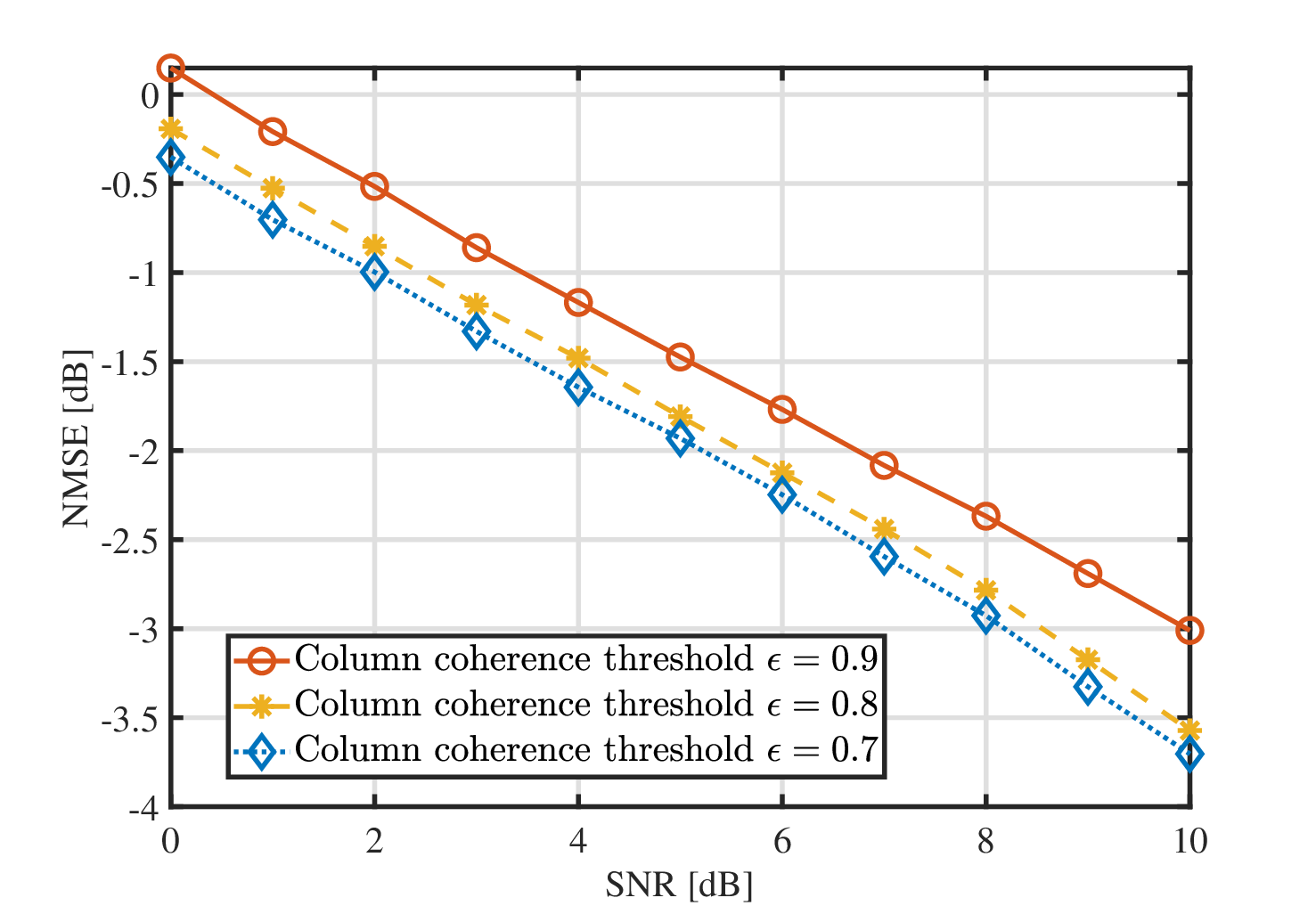} 
\caption{NMSE of different schemes against $\epsilon$.}
\label{column_change}
\end{figure}
In Fig. \ref{kappa_change}, we assess the NMSE performance of the proposed LCPD-SBL channel estimator as well as of the existing OMP and fast iterative shrinkage-thresholding algorithm (FISTA) \cite{BeckFISTA} against the control parameter $\kappa$. The SNR is set to $\SI{5}{dB}$, and the distance between the SIM and UEs (scatterers) follows $\mathcal{U}\left (r_{F}, \SI{10}{m} \right )$. As shown in Fig. \ref{kappa_change}, as the parameter $\kappa$ increases, the NMSE of both the proposed LCPD-SBL algorithm and the OMP algorithm (based on the polar-domain transform matrix) increases. Nonetheless, the LCPD-SBL algorithm consistently outperforms both the OMP and FISTA. Moreover, as $\kappa$ increases, the NMSE performance gap between the OMP algorithms based on the polar domain and the angular domain transform matrices is narrowed. This is because an increase in $\kappa$ reduces the number of sampling points $N_G$, leading to a decrease in grid resolution of the spatial sampling points. \par
Figure \ref{path_change} evaluates the estimation performance against the number of propagation paths $Q$. It should be noted that as the number of propagation paths increases, the NMSE performance of all algorithms deteriorates. This is because a larger number of propagation paths leads to more non-zero elements of the wireless channel in the sparse domain, thus degrading the performance of the CS-based channel estimators. However, the proposed LCPD-SBL algorithm still provides a $\SI{4}{dB}$ gain compared to both the OMP and FISTA algorithms. Moreover, in Fig. \ref{blocks_change}, as the number of pilot blocks increases, all algorithms' NMSE performance improves, which is due to the increase in the compression ratio from $1/16$ to $1/2$. In this context, the proposed LCPD-SBL algorithm demonstrates a more significant advantage in improving the NMSE performance.\par

Finally, to verify the benefits of the reduced column coherence of the polar-domain transform matrix on the channel recovery accuracy, we assess the NMSE performance attained by the LCPD-SBL algorithm in Fig. \ref{column_change}, with the polar-domain transform matrix designed by different column coherence thresholds. The UEs (scatterers) are uniformly distributed at a distance of ($\SI{0}{m}$, $\SI{10}{m}$) to contain all the distance sampling points. Figure \ref{column_change} illustrates that the polar-domain transform matrix constructed by using a lower column coherence threshold can enhance the ability of the CS framework to discriminate the paths' support set, thus improving the proposed LCPD-SBL algorithm's NMSE performance.

\section{Conclusions}
\label{Conclusion}
In this paper, we addressed the underdetermined near-field CE problem in SIM-based mmWave communication systems. Specifically, we designed a novel polar-domain transform matrix for UPAs and provided the polar-domain representation of the channel to transform the near-field CE problem into a sparse recovery task of the paths' support set and complex gains. Based on the near-field channel sparsity in the polar domain, we proposed an LCPD-SBL algorithm for estimating near-field channels, which can be up to $4\times$ faster than existing SBL methods. Simulation results indicated that, in the near-field region, the algorithms based on the proposed polar-domain transform matrix outperform those based on the angular-domain transform matrix. Moreover, the proposed LCPD-SBL algorithm also performed well in the far-field region.
\appendices
\section{Proof of Lemma 1}
\label{sam_angular}
{For two different elevation angles $\nu_{p}$ and $\nu_{q}$, we can approximate the column coherence as \cite{demir2023new}
\begin{equation}
\label{f3}
\begin{aligned}
 \bar{f}\left ( \nu_{p},\nu_{q},\iota,\iota,r,r \right ) &\approx \frac{1}{N}\left | \sum_{n=1}^{N_{V}}e^{-j\frac{2\pi \left ( n-\frac{N_{V}+1}{2} \right )\Delta}{\lambda }\left ( \nu _{p}-\nu _{q} \right ) } \right | \\
&\overset{\left(a\right)}{=}\frac{1}{N}\left | \frac{\text{sin}\left ( \frac{N_{V}\Delta\pi }{\lambda }\left ( \nu _{p}- \nu _{q}\right ) \right ) }{\text{sin}\left ( \frac{\Delta \pi}{\lambda }\left ( \nu _{p}- \nu _{q}\right ) \right )} \right | ,
\end{aligned}
\end{equation}
where $\overset{\left(a\right)}{=}$ is derived by $\frac{1-e^{-jN_{V}\Gamma}}{1-e^{-j\Gamma}}=e^{-j\frac{\left (N_{V}-1 \right )\Gamma }{2} }\frac{\text{sin}\left ( \frac{N_{V}\Gamma}{2} \right ) }{\text{sin}\left ( \frac{\Gamma}{2} \right )}$, and $\Gamma=\frac{2\pi\Delta\left ( \nu_{p}-\nu_{q} \right ) }{\lambda}$. To make the column coherence asymptotically approach zero, $\nu=\text{sin}\left ( \gamma \right ) $ should satisfy (\ref{nu}). Similarly, for the case with $\nu_{p}= \nu_{q} $, $r_{p}=r_{q}$, $\iota_{p}\ne\iota_{q}$, we can approximate the column coherence as 
\begin{equation}
\label{f4}
\begin{aligned}
 \bar{f}\left ( \nu,\nu,\iota_{p},\iota_{q},r,r \right ) &\approx \frac{1}{N}\left | \sum_{n=1}^{N_{H}}e^{-j\frac{2\pi \left ( n-\frac{N_{H}+1}{2} \right )\Delta}{\lambda }\left ( \iota _{p}-\iota _{q} \right ) } \right | \\
&\overset{\left(b\right)}{=}\frac{1}{N}\left | \frac{\text{sin}\left ( \frac{N_{H}\Delta\pi }{\lambda }\left ( \iota _{p}- \iota _{q}\right ) \right ) }{\text{sin}\left ( \frac{\Delta \pi}{\lambda }\left ( \iota _{p}- \iota _{q}\right ) \right )} \right | ,
\end{aligned}
\end{equation}
where $\overset{\left(b\right)}{=}$ is derived by $\frac{1-e^{-jN_{H}\kappa }}{1-e^{-j\kappa }}=e^{-j\frac{\left (N_{H}-1 \right )\kappa }{2} }\frac{\text{sin}\left ( \frac{N_{H}\kappa }{2} \right ) }{\text{sin}\left ( \frac{\kappa }{2} \right )}$, $\kappa =\frac{2\pi\Delta\left ( \iota_{p}-\iota_{q} \right ) }{\lambda}$. To make the column coherence asymptotically approach zero, $\iota = \text{cos}\left ( \gamma \right ) \text{sin}\left ( \vartheta \right ) $ should satisfy (\ref{Xi}). Finally, we consider the case with $r_{p}=r_{q}$, $\nu_{p}\ne\nu_{q} $, $\iota_{p}\ne\iota_{q}$, and approximate the column coherence as
\begin{equation}
\begin{aligned}
 &\bar{f}\left ( \nu_{p},\nu_{q},\iota_{p},\iota_{q},r,r \right )\\ \approx 
 &\left | \frac{\text{sin}\left ( \frac{N_{H}\Delta\pi }{\lambda }\left ( \iota _{p}- \iota _{q}\right ) \right ) }{N\text{sin}\left ( \frac{\Delta \pi}{\lambda }\left ( \iota _{p}- \iota _{q}\right ) \right )} \right | \! \times \! \left | \frac{\text{sin}\left ( \frac{N_{V}\Delta\pi }{\lambda }\left ( \nu _{p}- \nu _{q}\right ) \right ) }{\text{sin}\left ( \frac{\Delta \pi}{\lambda }\left ( \nu _{p}- \nu _{q}\right ) \right )} \right | .
\end{aligned}
\end{equation}
It can be inferred that when $\nu_{p}$, $\nu_{q}$ satisfy (\ref{nu}) or $\iota_{p}$, $\iota_{q}$ satisfy (\ref{Xi}), the column coherence becomes $0$. Therefore, if we sample the angular from all possible $\nu$ and $\iota$ in (\ref{nu}), (\ref{Xi}), and make sure $\iota^2+\nu^2=1-\text{cos}^2\left (\gamma \right ) \text{cos}^2\left (\vartheta \right ) \le 1$ \cite{demir2023new}, the column coherence of the above three cases can be effectively eliminated, which concludes the proof.
\section{Proof of Lemma 2}
\label{sam_distance}
The column coherence of two array response vectors in the same direction can be approximated as \cite{cui2022channel}
\begin{equation}
\label{f6}
\begin{aligned}
 &\bar{f}\left ( \nu,\nu,\iota,\iota,r_{p},r_{q} \right )\\&\overset{\left(c\right)}{\approx}\left |\int_{-\frac{1}{2} }^{\frac{1}{2}}e^{j\frac{\pi N_{H}^2\Delta^2\left ( \frac{1}{r_{p}}- \frac{1}{r_{q}} \right ) }{\lambda }t^2 }dt \right |\times \left |\int_{-\frac{1}{2} }^{\frac{1}{2}}e^{j\frac{\pi N_{V}^2\Delta^2\left ( \frac{1}{r_{p}}- \frac{1}{r_{q}} \right ) }{\lambda }t^2 }dt \right |\\ 
&=\left | G\left ( \varpi_{H} \right ) \right | \times \left | G\left ( \varpi_{V} \right ) \right |,
\end{aligned}
\end{equation}
where $\overset{\left(c\right)}{\approx}$ is derived by the definition of a definite integral, $\varpi_{H} =\sqrt{\frac{N_{H}^2\Delta^2 \left | \frac{1}{r_{p}}- \frac{1}{r_{q}} \right | }{2\lambda} }$, and $\varpi_{V} =\sqrt{\frac{N_{V}^2\Delta^2 \left | \frac{1}{r_{p}}- \frac{1}{r_{q}} \right | }{2\lambda} }$. The function $G\left ( \varpi\right ) = \frac{C\left ( \varpi \right )+jS\left ( \varpi \right ) }{\varpi } $ is composed of two Fresnel integrals, where $C\left ( \varpi \right ) =\int_{0}^{\varpi } \text{cos}\left ( \frac{\pi t^2}{2} \right )dt$, $S\left ( \varpi \right ) =\int_{0}^{\varpi } \text{sin}\left ( \frac{\pi t^2}{2} \right )dt $ \cite{sherman1962properties}. By assuming $\eta_{H}=\frac{\pi N_{H}^2\Delta^2\left ( \frac{1}{r_{p}} - \frac{1}{r_{q}}\right ) }{\lambda }$ and $\eta_{V}=\frac{\pi N_{V}^2\Delta^2\left ( \frac{1}{r_{p}} - \frac{1}{r_{q}}\right ) }{\lambda } $, then the column coherence considering $\iota_{p}\ne \iota_{q}$, $\nu_{p}\ne\nu_{q} $, and $r_{p}\ne r_{q}$ can be approximated as 
\begin{equation}
 \begin{aligned}
 &\bar{f}\left ( \nu_{p},\nu_{q},\iota_{p},\iota_{q},r_{p},r_{q} \right )
 \overset{\left(d\right)}{\approx} \left | e^{-j\frac{\pi^2i_{\iota}^2}{\eta_{H}} } \! \right | \! \left |\int_{\frac{-1}{2} \! - \! \frac{\pi i_{\iota}}{\eta_{H}} }^{\frac{1}{2} \! - \! \frac{\pi i_{\iota}}{\eta_{H}}}e^{j\eta_{H}u^2} du\right | \\ &\times
 \left | e^{-j\frac{\pi^2i_{\nu}^2}{\eta_{V}} } \right | \left |\int_{\frac{-1}{2} \! - \! \frac{\pi i_{\nu}}{\eta_{V}} }^{\frac{1}{2} \! - \! \frac{\pi i_{\nu}}{\eta_{V}}}e^{j\eta_{V}u^2} du\right |,
 \end{aligned}
\end{equation}
where $\overset{\left(d\right)}{\approx}$ is derived by the definition of a definite integral. By using the sampling method of angular parameters of (\ref{nu}), (\ref{Xi}), we obtain the terms $\iota_{p}-\iota_{q}=\frac{i_{\iota}\lambda }{N_{H}\Delta},\left | i_{\iota} \right | < \left \lceil \frac{2N_{H}\Delta}{\lambda } \right \rceil -1 $, and $\nu_{p}-\nu_{q}=\frac{i_{\nu}\lambda }{N_{V}\Delta},\left | i_{\nu} \right | < \left \lceil \frac{2N_{V}\Delta}{\lambda } \right \rceil -1 $. As $N_{H}\to \infty$, and $N_{V}\to \infty$, we have $\frac{\pi i_{\iota} }{\eta_{H}} \to 0$, and $\frac{\pi i_{\nu}}{\eta_{V}} \to 0$. Therefore, we can approximate the column coherence as
\begin{equation}
\label{f9}
\begin{aligned}
 \bar{f}\left ( \nu_{p},\nu_{q},\iota_{p},\iota_{q},r_{p},r_{q} \right )& \! \approx \! \left |\int_{-\frac{1}{2} }^{\frac{1}{2}}e^{j\eta_{H}u^2 }du \right | \! \times \! \left |\int_{-\frac{1}{2} }^{\frac{1}{2}}e^{j\eta_{V}u^2 }du \right |\\ 
&=\left | G\left ( \varpi_{H} \right ) \right | \times \left | G\left ( \varpi_{V} \right ) \right |.
\end{aligned}
\end{equation}
For simplicity, we define a function 
\begin{equation}
 \stackrel\frown{G} \left ( \kappa \right ) \overset{\Delta}{=} \left | G\left ( \varpi_{H}\right ) \right |\times \left | G\left ( \varpi_{V}\right ) \right |,
\end{equation}
where $\kappa =\left | \frac{1}{r_{p}} -\frac{1}{r_{q}} \right | $ is the control parameter. For a given column coherence threshold $\epsilon$, we can obtain a minimum $\kappa_{\epsilon}$ to ensure
$\stackrel\frown{G} \left ( \kappa_{\epsilon} \right )\le \epsilon$. Assuming that an initial sampling distance point $r_{0} \to \infty$, we can obtain (\ref{r}), which ensures that the column coherence of two array response vectors sampled at different distances is less than or equal to $\epsilon$.
\ifCLASSOPTIONcaptionsoff
 \newpage
\fi
\bibliographystyle{IEEEtran}
\bibliography{IEEEabrv,Ref}

\begin{IEEEbiography}[{\includegraphics[width=1in,height=1.25in,clip,keepaspectratio]{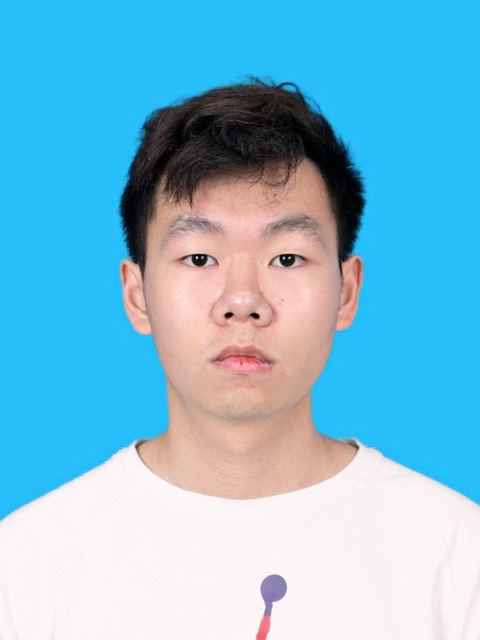}}]{Xianghao Yao}
received the B.E. degree from Jiangxi Normal University, Nanchang, China, in 2022, and the M.S. degree in information and communication engineering from the University of Electronic Science and Technology of China (UESTC), Chengdu, China, in 2025. His research interests include stacked intelligent metasurfaces, channel estimation, and compressed sensing. He was a recipient of the Talent Development Fund of Ouyang Minggao Academician Workstation in 2024.
\end{IEEEbiography}

\begin{IEEEbiography}[{\includegraphics[width=1in,height=1.25in,clip,keepaspectratio]{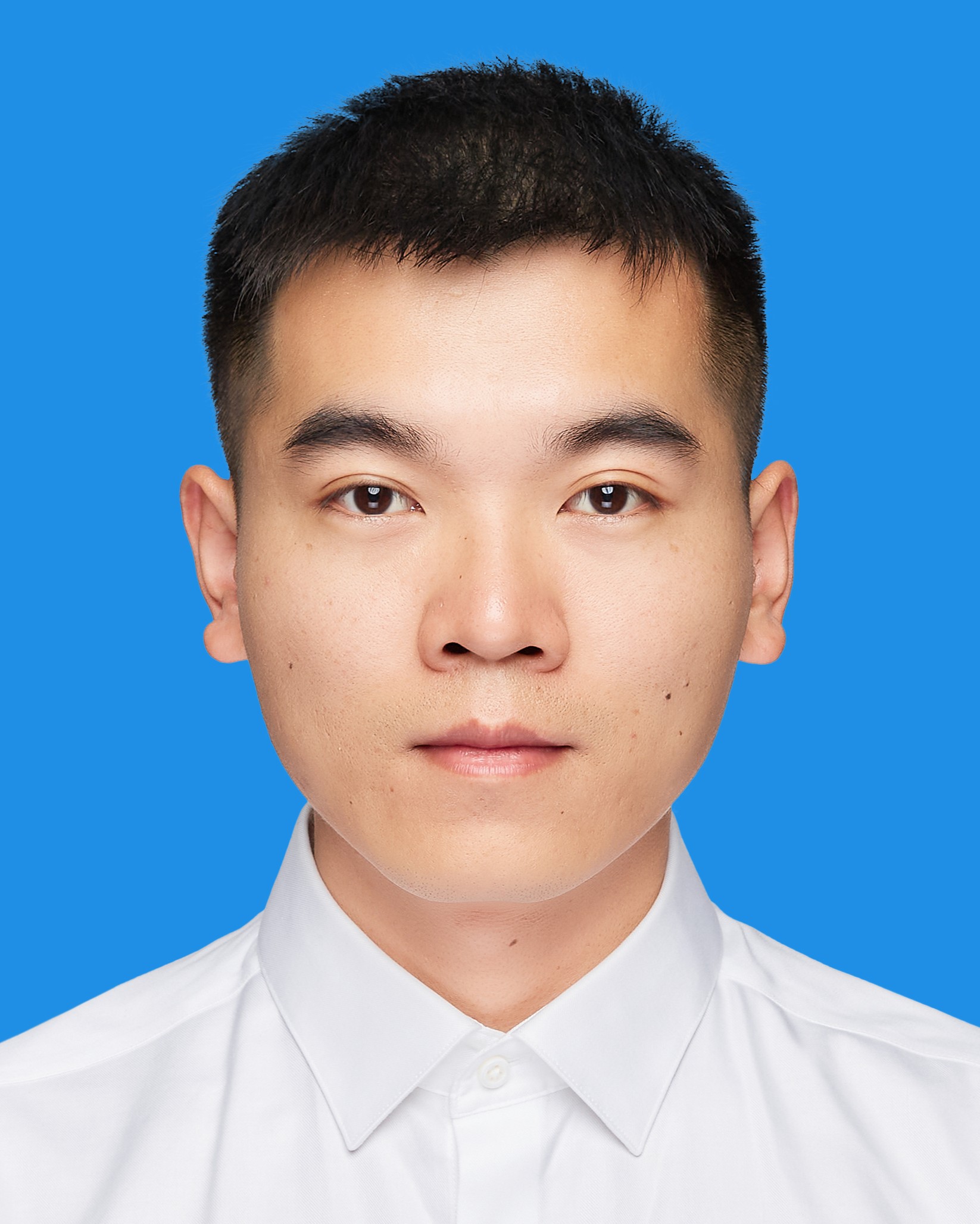}}]{Jiancheng An}
(Senior Member, IEEE) received the B.S. degree in Electronics and Information Engineering and the Ph.D. degree in Information and Communication Engineering from the University of Electronic Science and Technology of China (UESTC), Chengdu, China, in 2016 and 2021, respectively. From 2019 to 2020, he was a Visiting Scholar with the Next-Generation Wireless Group, University of Southampton, Southampton, U.K. From 2021 to 2023, he was a Post-Doctoral Research Fellow with the Engineering Product Development (EPD) Pillar, Singapore University of Technology and Design (SUTD), Singapore. From 2023 to 2026, he was a Research Fellow with the School of Electrical and Electronics Engineering, Nanyang Technological University (NTU), Singapore. He is currently a Professor with the School of Electronic Science and Engineering, UESTC, Chengdu, China.

Dr. An received the IEEE International Conference on Communications (ICC) 2023 Best Paper Award. He is the co-inventor of six patents and has published over 100 research papers in peer-reviewed international journals and conferences. His research interests include stacked intelligent metasurfaces (SIM), flexible intelligent metasurfaces (FIM), and electromagnetic neural networks (EMNN). Dr. An serves as an Editor for IEEE Transactions on Communications, IEEE Open Journal of the Communications Society, and IEEE Wireless Communications Letters. He is also the Lead Guest Editor for the Special Issue on “Stacked Intelligent Metasurface-Empowered Advanced Signal Processing Paradigm for 6G and Beyond” in IEEE Wireless Communications.
\end{IEEEbiography}

\begin{IEEEbiography}[{\includegraphics[width=1in,height=1.25in,clip,keepaspectratio]{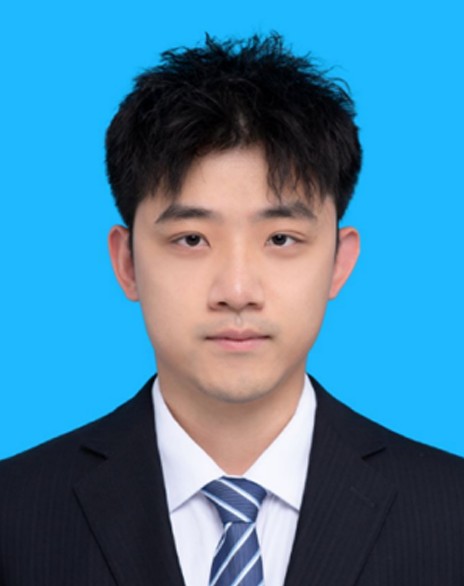}}]{Enyu Shi}
(Member, IEEE) received the B.S. and Ph.D. degrees in communication engineering from Beijing Jiaotong University, Beijing, China, in 2019 and 2025, respectively. He was a Visiting Student with the School of Electrical and Electronic Engineering, Nanyang Technological University, Singapore, from 2024 to 2025. He is currently a Post-Doctoral Research Fellow with Beijing Jiaotong University. His research interests include cell-free massive MIMO, RIS/SIM, signal processing, and artificial intelligence of wireless systems. He received the Best Paper Award at IEEE WCSP 2024.
\end{IEEEbiography}

\begin{IEEEbiography}[{\includegraphics[width=1in,height=1.25in,clip,keepaspectratio]{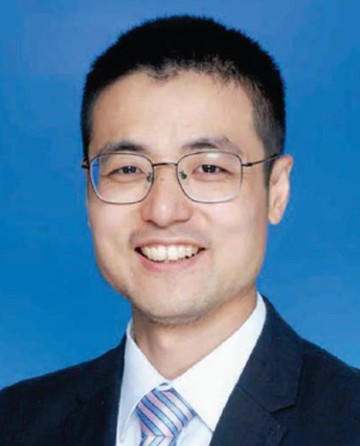}}]{Jiayi Zhang}
(Senior Member, IEEE) received the B.Sc. and Ph.D. degrees in communication engineering from Beijing Jiaotong University, China, in 2007 and 2014, respectively.

From 2012 to 2013, he was a Visiting Scholar with the Wireless Group, University of Southampton, U.K. From 2014 to 2015, he was also a Humboldt Research Fellow with the Institute for Digital Communications, Friedrich-Alexander University of Erlangen–Nuremberg (FAU), Germany. From 2014 to 2016, he was a Post-Doctoral Research Associate with the Department of Electronic Engineering, Tsinghua University, China. Since 2016, he has been a Professor with the School of Electronic and Information Engineering, Beijing Jiaotong University. His current research interests include cell-free massive MIMO, reconfigurable intelligent surface (RIS), XL-MIMO, near-field communications, and applied mathematics. He received the Best Paper Award from IEEE OPEN JOURNAL OF THE COMPUTER SOCIETY in 2023, the Best Paper Awards at IEEE ICC 2023 and WCSP 2024, the URSI Young Scientist Award in 2020, and the IEEE ComSoc Asia–Pacific Outstanding Young Researcher Award in 2020. He was the Lead Guest Editor of the Special Issue on “Multiple Antenna Technologies for Beyond 5G” of IEEE JOURNAL ON SELECTED AREAS IN COMMUNICATIONS, the Lead Guest Editor of the Special Issue on “Semantic Communications for the Metaverse” of IEEE WIRELESS COMMUNICATIONS, an Editor of IEEE COMMUNICATIONS LETTERS from 2016 to 2021, and an Editor of IEEE TRANSACTIONS ON COMMUNICATIONS from 2019 to 2024. He serves as an Associate Editor for IEEE TRANSACTIONS ON WIRELESS COMMUNICATIONS.
\end{IEEEbiography}

\begin{IEEEbiography}[{\includegraphics[width=1in,height=1.25in,clip,keepaspectratio]{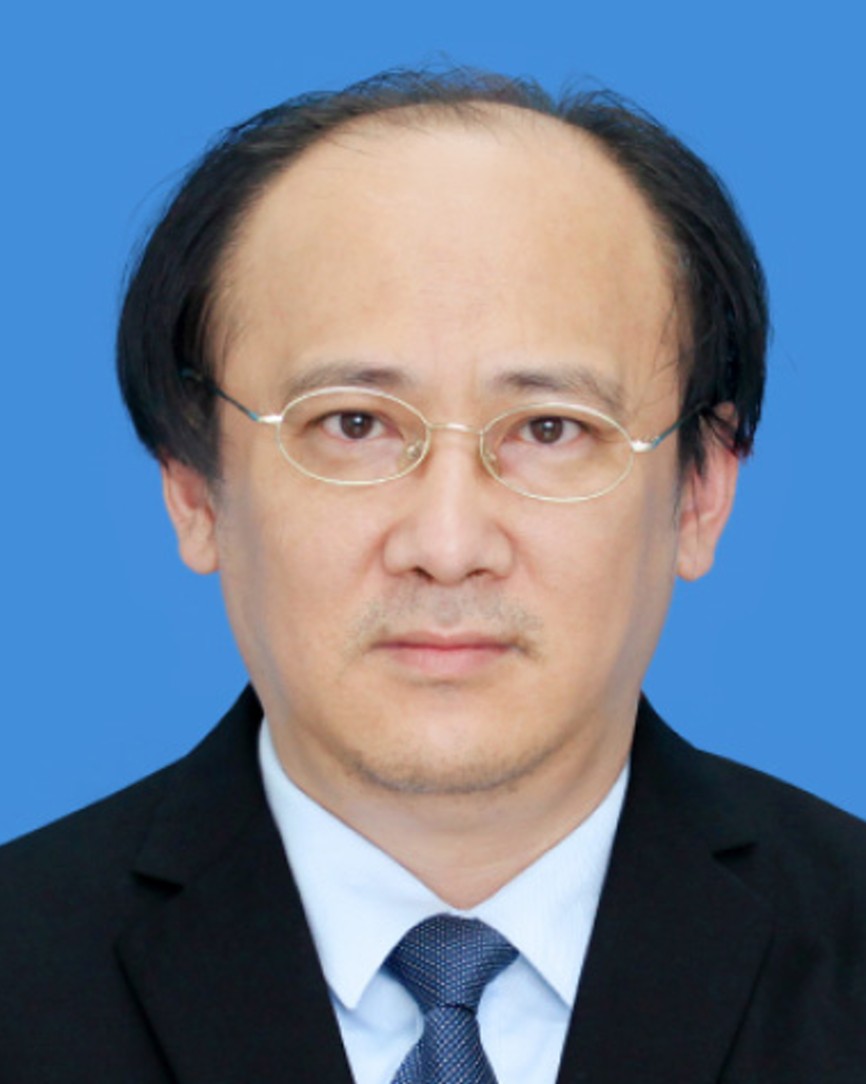}}]{Lu Gan}
(Member, IEEE) received the M.S. and Ph.D. degrees from the University of Electronic Science and Technology of China (UESTC), Chengdu, China, in 2002 and 2006, respectively. From September 2012 to September 2013, he was a Visiting Researcher with the University of Concordia, Montreal, QC, Canada. Since August 2014, he has been a Professor with UESTC. His research interests include signal detection and classification, array signal processing, compressive sensing, passive radar, and reconfigurable intelligent surfaces.
\end{IEEEbiography}

\begin{IEEEbiography}[{\includegraphics[width=1in,height=1.25in,clip,keepaspectratio]{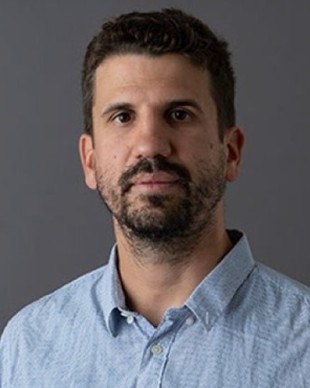}}]{Michail Matthaiou}
(Fellow, IEEE) received the Ph.D. degree from the University of Edinburgh, U.K., in 2008.

He is currently a Professor of Communications Engineering and Signal Processing and the Deputy Director of the Centre for Wireless Innovation (CWI), Queen’s University Belfast, U.K. He is also an Eminent Scholar at Kyung Hee University, Republic of Korea. He has held research/faculty positions at Munich University of Technology (TUM), Germany, and Chalmers University of Technology, Sweden. He holds the ERC Consolidator Grant BEATRICE (2021–2027), focused on the interface between information and electromagnetic theories. His research interests span signal processing for wireless communications, beyond massive MIMO, reflecting intelligent surfaces, mm wave/THz systems, and AI-empowered communications. He is an AAIA Fellow. He and his co-authors received the IEEE Communications Society (ComSoc) Leonard G. Abraham Prize in 2017. To date, he has received the prestigious 2023 Argo Network Innovation Award, the 2019 EURASIP Early Career Award, and the 2018/2019 Royal Academy of Engineering/The Leverhulme Trust Senior Research Fellowship. His team was also the Grand Winner of the 2019 Mobile World Congress Challenge. He was a recipient of the 2011 IEEE ComSoc Best Young Researcher Award for the Europe, Middle East, and Africa Region. He was a co-recipient of the 2006 IEEE Communications Chapter Project Prize for the best M.Sc. dissertation in the area of communications. He has co-authored papers that received best paper awards at the 2018 IEEE WCSP and 2014 IEEE ICC. In 2014, he received the Research Fund for International Young Scientists from the National Natural Science Foundation of China. He is the Editor-in-Chief of Physical Communication (Elsevier), a Senior Editor of IEEE WIRELESS COMMUNICATIONS LETTERS and IEEESignal Processing Magazine, an Area Editor of IEEE TRANSACTIONS ON COMMUNICATIONS, and the Editor-in-Large of IEEE OPEN JOURNAL OF THE COMMUNICATIONS SOCIETY.
\end{IEEEbiography}

\begin{IEEEbiography}[{\includegraphics[width=1in,height=1.25in,clip,keepaspectratio]{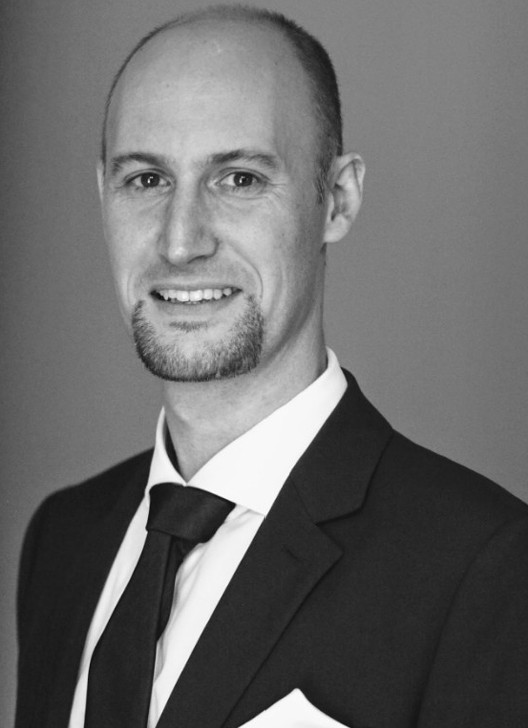}}]{Symeon Chatzinotas}
(Fellow, IEEE) is currently Full Professor / Chief Scientist I and Head of the research group SIGCOM in the Interdisciplinary Centre for Security, Reliability and Trust, University of Luxembourg. In parallel, he is an Adjunct Professor in the Department of Electronic Systems, Norwegian University of Science and Technology, an Eminent Scholar of the Kyung Hee University, Korea and a Collaborating Scholar of the Institute of Informatics \& Telecommunications, National Center for Scientific Research “Demokritos”.

In the past, he has been a Visiting Professor at EPFL, Switzerland and University of Parma, Italy and contributed in numerous R\&D projects for the Institute of Telematics and Informatics, Center of Research and Technology Hellas and Mobile Communications Research Group, Center of Communica tion Systems Research, University of Surrey. He has received the M.Eng. in Telecommunications from Aristotle University of Thessaloniki, Greece and the M.Sc. and Ph.D. in Electronic Engineering from University of Surrey, UK in 2003, 2006 and 2009 respectively. He has authored more than 800 technical papers in refereed international journals, conferences and scientific books and has received numerous awards and recognitions, including the IEEE Fellowship, IEEE Distinguished Contributions Award and IEEE Harry Rowe Mimno Award. He has served in the editorial board of npj Wireless Technology, IEEE Transactions on Communications, IEEE Open Journal of Vehicular Technology and the International Journal of Satellite Communications and Networking.
\end{IEEEbiography}

\begin{IEEEbiography}[{\includegraphics[width=1in,height=1.25in,clip,keepaspectratio]{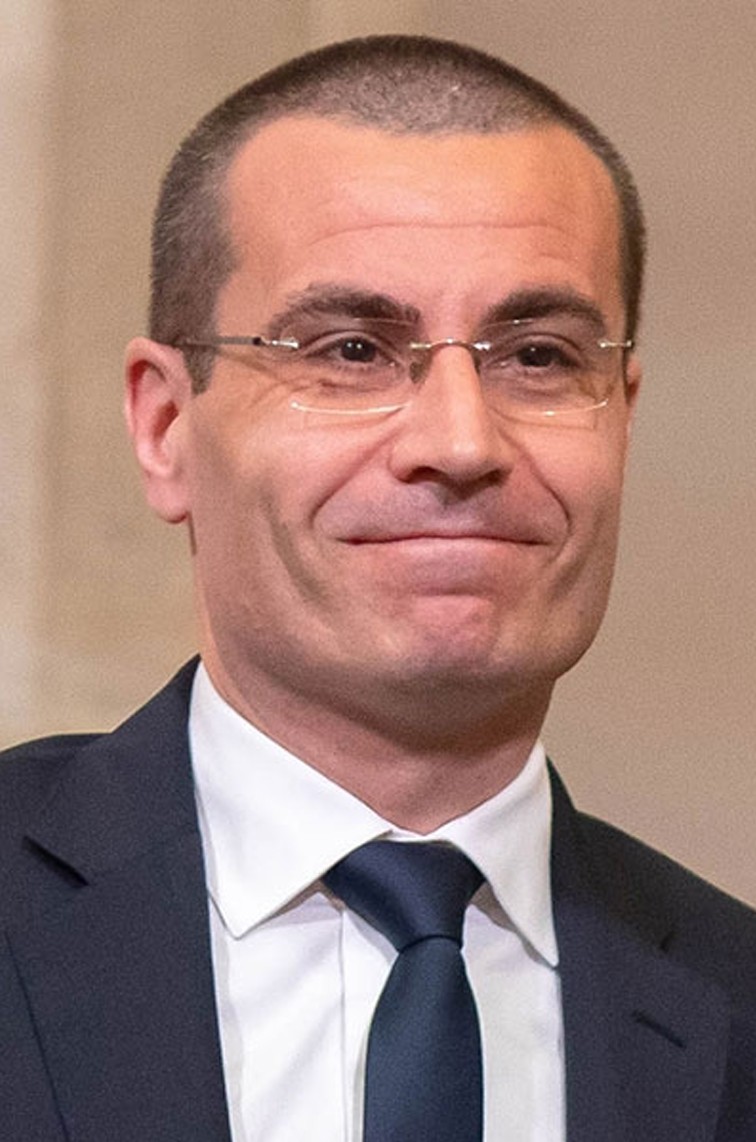}}]{Marco Di Renzo}
(Fellow, IEEE) received the Laurea (cum laude) and Ph.D. degrees in electrical engineering from the University of L’Aquila, Italy, in 2003 and 2007, respectively, and the Habilitation à Diriger des Recherches (Doctor of Science) degree from University Paris-Sud (currently Paris-Saclay University), France, in 2013. Currently, he is Chair Professor of Telecommunications Engineering, the Director of the Centre for Telecommunications Research, and the Head of the Telecommunications Group, Department of Engineering, King’s College London, London, United Kingdom. He is also a CNRS Research Director (Professor) with the Institute of Electronics and Digital Technologies (IETR) at CNRS-CentraleSup\'elec, Rennes, France. He was a France-Nokia Chair of Excellence in ICT at the University of Oulu (Finland), a Tan Chin Tuan Exchange Fellow in Engineering at Nanyang Technological University (Singapore), a Fulbright Fellow at The City University of New York (USA), a Nokia Foundation Visiting Professor at Aalto University (Finland), and a Royal Academy of Engineering Distinguished Visiting Fellow at Queen’s University Belfast (U.K.). He is a Fellow of the IEEE, IET, EURASIP, and AAIA; an Academician of AIIA; an Ordinary Member of the European Academy of Sciences and Arts, an Ordinary Member of the Academia Europaea, and an Ordinary Member of the Italian Academy of Technology and Engineering; an Ambassador of the European Association on Antennas and Propagation; and a Highly Cited Researcher. He has received several distinctions, including the Michel Monpetit Prize conferred by the French Academy of Sciences, the IEEE Communications Society Heinrich Hertz Award, and the IEEE Communications Society Marconi Prize Paper Award in Wireless Communications. Also, he is a principal investigator of an ERC Synergy grant on metasurface-based information processing. He served as the Editor-in-Chief of IEEE Communications Letters from 2019 to 2023, and as the Director of Journals and Chair of the Publications Misconduct Ad Hoc Committee of the IEEE Communications Society from 2024 to 2025. Currently, he sits on the IEEE-COMSOC Fellow Evaluation Standing Committee and on the Editorial Board of the Proceedings of the IEEE.
\end{IEEEbiography}

\end{document}